\documentclass[journal=jacsat,manuscript=article]{achemso}

\usepackage{chemformula} 
\usepackage[T1]{fontenc} 
\usepackage{graphicx}
\usepackage[version=3]{mhchem}
\usepackage{subcaption}
\usepackage{bm}
\usepackage{amssymb}
\usepackage{caption}
\usepackage{amsmath}
\usepackage[T1]{fontenc}
\usepackage[utf8]{inputenc}
\usepackage{amsbsy}
\usepackage[usetitle]{rsc}

\author{Amir Hossein Raffiee}
\affiliation{School of Mechanical Engineering, Purdue University, West Lafayette, IN 47907, USA}
\author{Arezoo M. Ardekani}
\affiliation{School of Mechanical Engineering, Purdue University, West Lafayette, IN 47907, USA}
\email{ardekani@purdue.edu}
\author{Sadegh Dabiri}
\affiliation{Department of Agricultural and Biological Engineering,
Purdue University, West Lafayette, IN 47907, USA}
\alsoaffiliation{School of Mechanical Engineering, Purdue University, West Lafayette, IN 47907, USA}

\title[An \textsf{achemso} demo]
  {Numerical investigation of elasto-inertial particle focusing patterns in viscoelastic microfluidic devices}


\begin{document}







\begin{abstract}
  Viscoelastic microfluidic devices are promising for various microscale procedures such as particle sorting and separation. 
  In this study, we perform three dimensional computational investigation of a particle in a viscoelastic channel flow by considering combined inertial and elastic effects. 
  We calculate equilibrium positions of the particle in the microchannel for a wide range of Reynolds and Weissenberg numbers, which are important in the design of microfluidic devices.  
  The results provide an insight into the motion of cells and particles and explain the findings of previous experiments. Furthermore, we  suggest new particle behaviors that have not been found before.  Ultimately, a phase diagram is provided to predict the particle dynamics under a wide range of inertial and elastic effects.
\end{abstract}

\section{Introduction}
Microfluidic devices have been widely used in biomedical devices. These devices have significantly enhanced therapeutic, diagnostic and many industrial procedures  by increasing accuracy and accelerating the processes\cite{gossett2010label,karimi2013hydrodynamic,khalesi2019numerical,yamashita2017microencapsulation,di2009inertial,castro2018portable,sadeghi2018coalescence}. Isolation and separation of rare cells from a heterogeneous population of cells is a critical process in early diagnosis of fatal diseases such as cancer\cite{van2011circulating} and malaria\cite{gascoyne2004microfluidic}. Furthermore, the isolated cells such as rare blood components can  be used for therapeutic purposes. For instance, platelet-rich plasma are used in transfusion\cite{sethu2006microfluidic,gossett2010label} or stem cells can be found in blood samples\cite{gossett2010label}. Enriching the cell population provides a platform to biologists to conduct physical and chemical analysis on cells\cite{choi2009microfluidic,hafezisefat2017experimental,el2006cells,park2018towards,almansoori2018parametric,sadri2017aggregation}.

In order to accomplish the desired tasks in aforementioned applications the precise control of particles is required\cite{del2017edge}. Hence, developing new techniques for the manipulation of particle trajectory  has been the subject of many studies in the past decade. Some of these techniques are designed based on the use of externally applied forces generated by electric\cite{pethig2010dielectrophoresis}, magnetic\cite{pamme2006magnetism} and acoustic\cite{friend2011microscale} fields. These methods offer high sample processing rates\cite{gossett2010label}, however, there are many factors that prevent them from widely being used in clinical applications. Mainly, these methods work based on biochemical labeling of the cells that may affect the cell function and properties\cite{karimi2013hydrodynamic,gossett2010label}. Furthermore, the cost and complexity of the process can also be considered as an important downside for these methods\cite{gossett2010label}. Hence, there is a growing interest in developing label-free techniques that introduce advantages such as accurate analysis, low sample use and low cost operations\cite{karimi2013hydrodynamic}. In this regard, inertial microfluidic devices are used to manipulate particle trajectories in microchannels. 

\citeauthor{segre1961radial}\cite{segre1961radial} first reported the transverse migration of particles due to inertial forces in a straight tube filled by a Newtonian fluid\cite{segre1961radial}. The hydrodynamic interaction between the flowing particles and the ambient fluid can be used to manipulate the trajectory of particles. This effect was studied theoretically by \cite{ho1974inertial} in which they considered the dynamics of a rigid sphere in a 2D Poiseuille flow and a simple shear flow and predicted the equilibrium locations of the particle in these flows. Furthermore, they showed that the final location of the particle is independent of releasing point in the microchannel. \cite{ishii1980lateral} also theoretically investigated the lateral migration of a solid particle in a circular tube. They studied the motion of a buoyant and neutrally buoyant particle in a Poiseuille flow and showed that their findings agree well with experiments. Many microfluidic setups are designed based on inertial migration of particles to control the location of targeted cells in microchannels, particularly for sorting and separation of the cells\cite{di2007continuous,d2017particle,choi2011lateral,isiksacan2018lab,isiksacan2017microfluidic,hur2010sheathless}. The efficiency of this method reduces in cases where the targeted cells are small or the flow rate is low\cite{karnis1966particle,lu2017particle,del2017edge}. Viscoelastic microfluics are developed to address this issue by replacing the Newtonian ambient fluid by a dilute polymer solution\cite{karnis1966particle}. The solute polymer chains get deformed in the induced flow field and exert an additional elastic force on the particle. The resulting force affects the particle migration along with the inertial force\cite{del2017edge,d2017particle}. Previous studies show that the direction and the magnitude of generated elastic force depends on the rheology of the polymer solution and the volumetric flow rate of the suspending fluid\cite{karnis1966particle,gauthier1971particle,leshansky2007tunable}.  

Transverse migration of a particle suspended in a viscoelastic fluid is caused by the lift force generated due to the interaction between ambient flow and the particle\cite{ho1976migration,chan1977note,brunn1980motion}. This lift force comprises an inertial lift force ($F_{in}$) and an elastic lift force ($F_{el}$). The inertial lift force can be decomposed into two forces: (i) shear-gradient lift force ($F_{s\_in}$) that arises from the non-uniform velocity profile across the channel and drives the particle away from the channel center and (ii) wall-induced lift force ($F_{w\_in}$) that is caused by the uneven distribution of the vorticity around the particle that leads to higher pressure in the gap between the wall and the particle and pushes the particle away from the wall\cite{leal1980particle,ho1974inertial,zeng2005wall,zeng2009forces}. These forces have been investigated extensively in the literature and there are experimental\cite{karnis1966flow,matas2004inertial}, numerical\cite{feng1994direct,trofa2015numerical,liu2016generalized,di2009particle} and analytical\cite{asmolov1999inertial,ho1974inertial,chan1977note,brunn1980motion} works proposing scaling relations for the total inertial force in square and rectangular microchannels. On the other hand, the elastic force acts on the particle due to non-uniform distribution of normal stress difference across the channel \cite{ho1976migration,huang1997direct}. In this phenomenon, the first normal stress difference ($N_1$) generates a stream-wise tension and the second normal stress difference ($N_2$) gives rise to a secondary flow in the cross section\cite{villone2013particle}. 
There are numerical\cite{wang2018numerical,villone2013particle,raffiee2017elasto,raffiee2019suspension,saadat2018immersed} and experimental\cite{del2017edge,seo2014lateral} studies investigating the particle migration in a viscoelastic fluid. Here, in this work we focus on illustrating the distribution of elastic and inertial forces acting on a particle  in a viscoelastic channel flow and calculating the equilibrium positions of the particle in the channel cross section.

In order to understand the mechanism of particle migration in a viscoelastic fluid, fully resolved 3D numerical simulations are conducted. In this work, we show the distribution of lift force acting on the particle in a viscoelastic fluid and investigate the influence of combined elastic and inertial forces on the particle behavior in a microchannel. Furthermore, the location of equilibrium points and their corresponding stability are determined for $1<Re<30$ and $0<Wi<3$ which is important for designing the microfluidic devices relying on viscoelastic effects. Our results explain various focusing pattern of particles observed in previous experimental works by scrutinizing the stability of equilibrium points. We also predict new behaviors that have not been discovered in the past studies. 

\section{Mathematical modeling}
\begin{figure}[h!]
    \centering
        \includegraphics[height=2.8in]{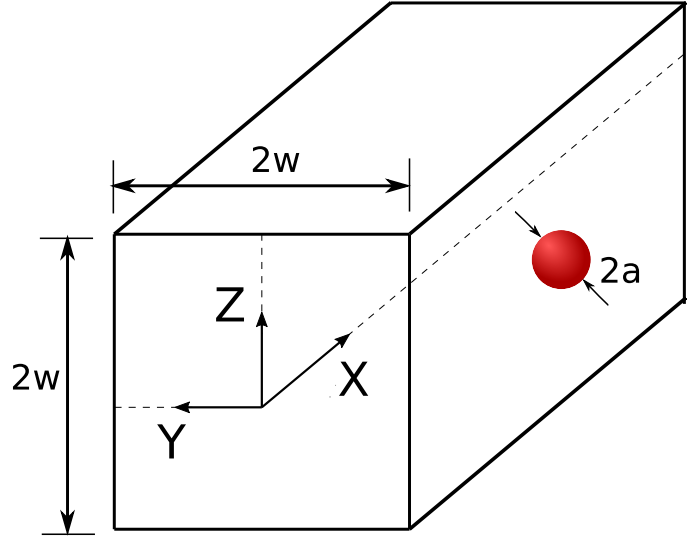}
\caption{Schematic of the problem setup }
    \label{T1}
\end{figure}

In this work, we study the dynamics of a solid spherical particle suspended in a straight, square microchannel. The particle is neutrally buoyant and the ambient fluid is viscoelastic. The radius of the sphere is represented by $a$ and the channel side and its length are $2w$ and $L$, respectively. In this problem, the origin of the reference frame is located at the channel center and $x$, $y$ and $z$ directions are aligned with the streamwise, horizontal and vertical directions, respectively, as illustrated in Fig. \ref{T1}. The particle is initially at rest and a constant pressure gradient is applied in the $x$ direction, driving the flow in the microchannel. In order to simulate the hydrodynamic interaction between the particle and the surrounding fluid, an incompressible Navier-Stokes equation is numerically solved  as follows:
\begin{equation}\label{eq:0}
\nabla .\textbf{u}=0,
\end{equation}
\begin{equation}\label{eq:1}
\frac{\partial (\rho \textbf{u})}{\partial t}+\nabla.(\rho \textbf{uu})=-\nabla p+\nabla .\mbox{\boldmath$\tau$}+\textbf{F},
\end{equation}
where $\rho$ is the fluid density, \textbf{u} is the velocity vector, $t$ denotes the time and $p$ and \mbox{\boldmath$\tau$} represent the pressure and stress tensor, respectively. The particle density is assumed to be the same as the fluid density. A Distributed Lagrangian Multiplier (DLM) method is used to simulate the solid particle motion in the fluid. A forcing term $\textbf{F}$ is added in eq. \ref{eq:1} to enforce the rigid body motion of the  particle. The details of DLM method can be found in \cite{ardekani2008collision} \cite{ardekani2008collision} The viscoelastic properties of the fluid can be modeled by splitting the stress tensor into two parts: (i) the contribution from solvent \mbox{\boldmath$\tau_s$} and (ii) that of polymer \mbox{\boldmath$\tau_p$}. Hence, the total stress tensor can be written as:
\begin{equation}\label{eq:2}
\mbox{\boldmath$\tau$}=\mbox{\boldmath$\tau_s$}+\mbox{\boldmath$\tau_p$}.
\end{equation}  
The Newtonian viscous stress is described as $\mbox{\boldmath$\tau_s$}=\mu_s (\nabla \textbf{u} +\nabla \textbf{u}^{T})$, where  $\mu_s$ represents the solvent viscosity. The Giesekus constitutive equation \cite{giesekus1982simple} is used to model the viscoelastic behavior of the fluid. This model captures the shear-thinning behavior and constrained elongation of polymer chains in the fluid\cite{li2015dynamics}. According to this model, the polymeric stress tensor is governed by:
\begin{equation}\label{eq:3}
\lambda \overset{\bigtriangledown} {\mbox{\boldmath$\tau_p$}} +\mbox{\boldmath$\tau_p$}+\frac{\alpha \lambda}{\mu_p}\mbox{\boldmath$\tau_p$}.\mbox{\boldmath$\tau_p$}=\mu_p (\nabla \textbf{u} +\nabla \textbf{u}^{T}),
\end{equation}
\begin{equation}\label{eq:4}
\overset{\bigtriangledown}{\mbox{\boldmath$\tau_p$}}\equiv \frac{\partial \mbox{\boldmath$\tau_p$}}{\partial t}+\textbf{u}.\nabla \mbox{\boldmath$\tau_p$}-\nabla \textbf{u} \mbox{\boldmath$\tau_p$}-\mbox{\boldmath$\tau_p$}\nabla \textbf{u}^T.
\end{equation}
Here $\lambda$ represents the fluid relaxation time and $\mu_p$ and $\alpha$ denote the polymeric viscosity and the mobility factor.  The following equations are used to calculate the total, elastic and inertial forces acting on the particle.
\begin{equation}\label{eq:5}
\textbf{F}_{total} = -\oint_V \textbf{F} \,dv,
\end{equation}
\begin{equation}\label{eq:6}
\textbf{F}_{el} = \int \mbox{\boldmath$\tau_p$}.\textbf{n} \,ds,
\end{equation}
\begin{equation}\label{eq:7}
\textbf{F}_{in} = \textbf{F}_{total} - \textbf{F}_{el}, 
\end{equation}
where $\textbf{n}$ represents the unit vector normal to the particle surface and $ds$ and $dv$ denote differential elements of particle surface and volume, respectively.
In this problem, the no-slip boundary condition is applied at the walls of the channel. These boundaries are normal to y and z directions. The periodic boundary condition is applied at the inlet and outlet of the computational domain. These boundaries are normal to the x direction. 
The microchannel length is set to $L=20a$ to ensure the channel is long enough and the particle does not interact with its periodic image. The details of the numerical methods used in this work and their validations are reported in our previous works\cite{raffiee2017deformation,raffiee2017elasto,raffiee2019suspension,li2015dynamics}.
We use $w$ as the characteristic length scale and $U_0$ as the characteristic velocity scale (where $U_0$ is the undisturbed flow velocity at the channel center filled with a Newtonian fluid). Accordingly, the governing dimensionless numbers can be defined as: (i) $Re=\frac{\rho U_0 2w}{\mu}$,  representing the ratio of inertial to the viscous forces in which the total viscosity is defined as $\mu=\mu_s+\mu_p$, (ii) the Weissenberg number $Wi=\frac{\lambda U_0 }{w}$,  representing the ratio of elastic to the viscous forces, (iii) $\beta=\frac{\mu_{p}}{\mu}$, representing the ratio of polymer viscosity to the total viscosity and (iv) blockage ratio $\frac{a}{w}$, describing the finite size of the flowing particle. In this work, $\beta$, $\frac{a}{w}$ and $\alpha$ are set to $0.5$, $0.3$ and $0.2$, respectively, unless otherwise stated. The particle has the same density as that of the fluid. It should be noted that the initial velocity of the fluid is set to the undisturbed flow velocity in the absence of the particle. In order to find the lift force distribution experienced by the particle across the channel cross section, the lateral position of the particle is fixed at the location where the lift force should be calculated. Consequently, the particle only travels along a line parallel to the $x$ direction and it rotates freely around all directions. The particle is consequently not force-free and the elastic and inertial forces balance with the transverse force required to fix the transverse location of the particle. Hence, the dynamics of this particle is not the same as a force-free particle dynamics due to non-linearity of the problem, but the equilibrium positions and their instability are the same as the ones for a force-free particle. The particle is released with a zero initial velocity in the microchannel and the simulation continues until the ambient flow field reaches a steady state and then  the lift force is calculated. This method is used to determine the particle's equilibrium locations and their corresponding stability across the microchannel in a Newtonian fluid\cite{prohm2014feedback,di2009particle,liu2015inertial,gossett2012inertial,yang2005migration,liu2016generalized}. The domain and grid size independency tests and the validation of the employed method are shown in the supplementary material.  

\section{Results and discussion}
The migration of particles in Newtonian fluids has been extensively investigated by many researchers. According to the previous numerical and experimental studies\cite{segre1961radial,di2009inertial}, the particles released in microchannel with a circular cross-section migrate toward an annulus ring with a radius of $\sim0.6R$, where $R$ represents the radius of microchannel. However, the particle dynamics changes in a square microchannel due to the reduced level of symmetry in the flow field\cite{li2015dynamics,di2009inertial,seo2014lateral}. In this case, the particles focus at four discrete points along the main axes of the microchannel. Considering the flow structure, there are nine equilibrium points, where the lateral force becomes zero, among which only four are stable. The rest exhibit unstable behavior, meaning the particles migrate away upon any disturbance in the flow\cite{prohm2014feedback,di2009particle,di2009inertial}. This phenomenon is further explained in details in the following section. In this work, we investigate the stability and location of equilibrium points for the particles suspended in a viscoelastic fluid and we show the focusing patterns of particles for a wide range of $Re$ and $Wi$ numbers. The variety observed in the focusing patterns of particles indicates the promising effect of addition of polymers in microfluidic devices and we show that this method can be used for different applications without applying any changes to the  geometry of the microchannel.         

\subsection{Migration in a low inertial regime}
\begin{figure*}[ht!]
    \centering
    \begin{subfigure}[t]{.5\linewidth}
        \centering
        \includegraphics[height=2.4in]{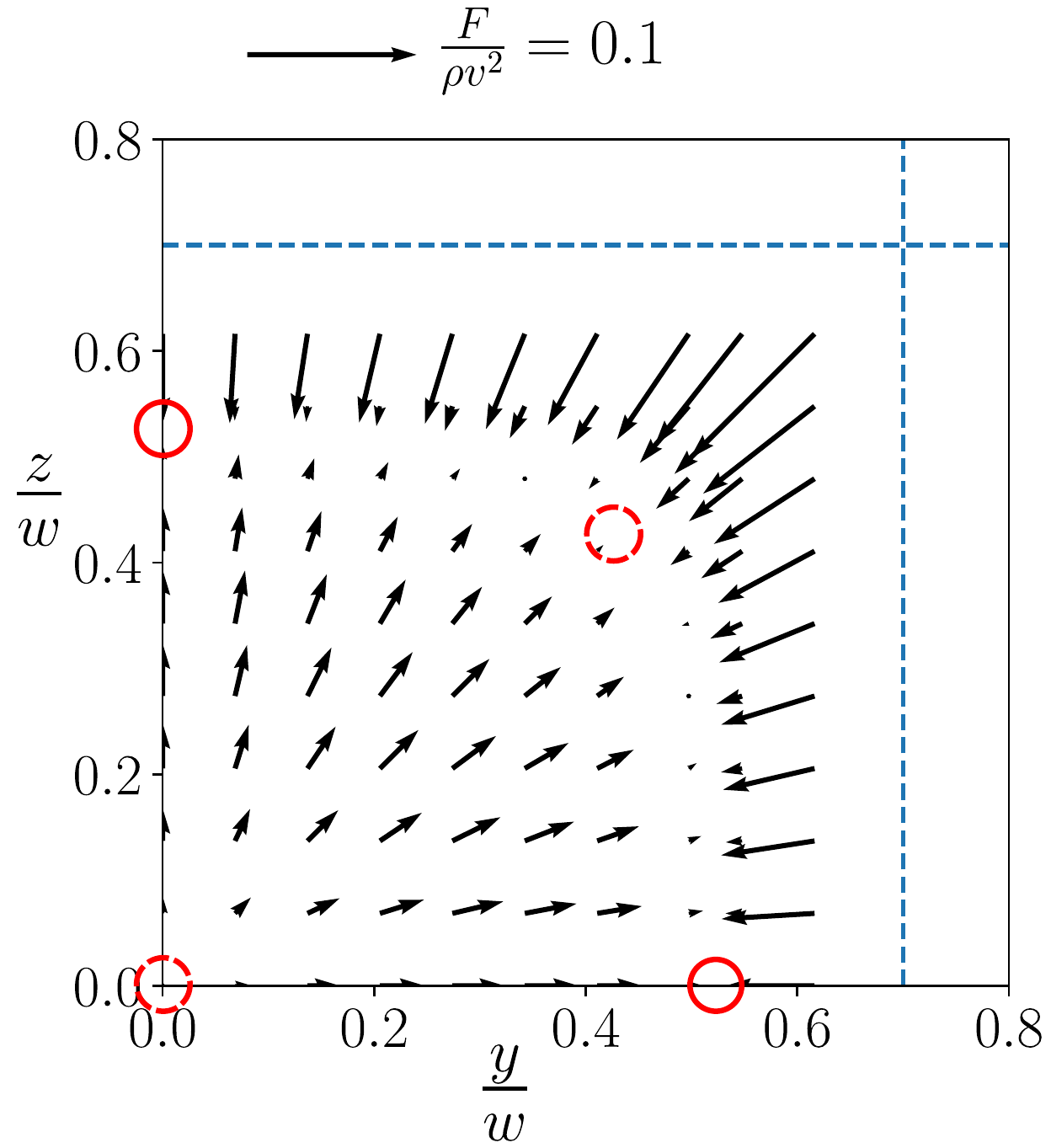}
        \caption{}
    \end{subfigure}%
    ~
    \begin{subfigure}[b]{.5\linewidth}
        \centering
        \includegraphics[height=1.9in]{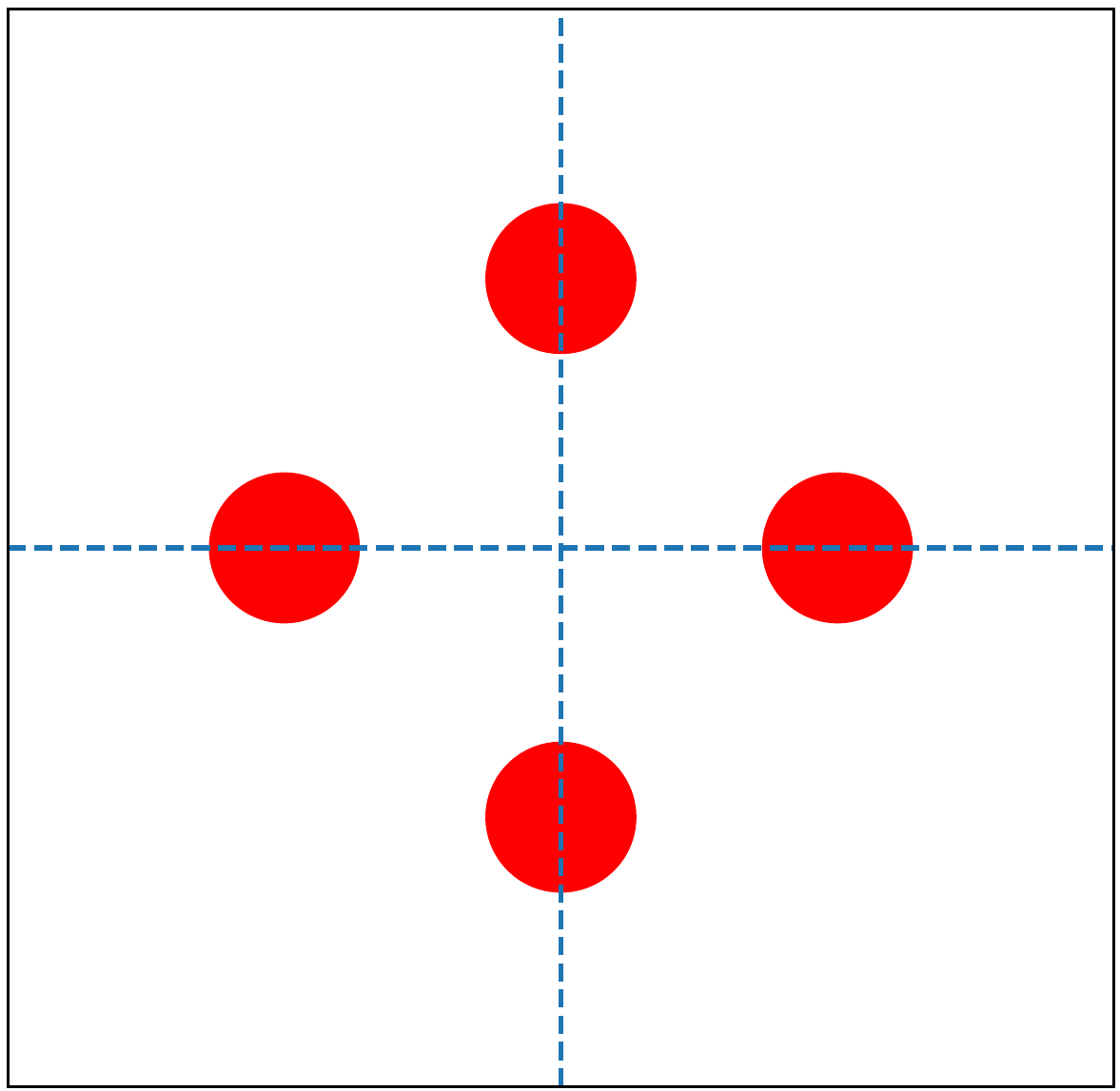}
        \caption{}
    \end{subfigure}
    
     \begin{subfigure}[t]{.5\linewidth}
        \centering
        \includegraphics[height=2.2in]{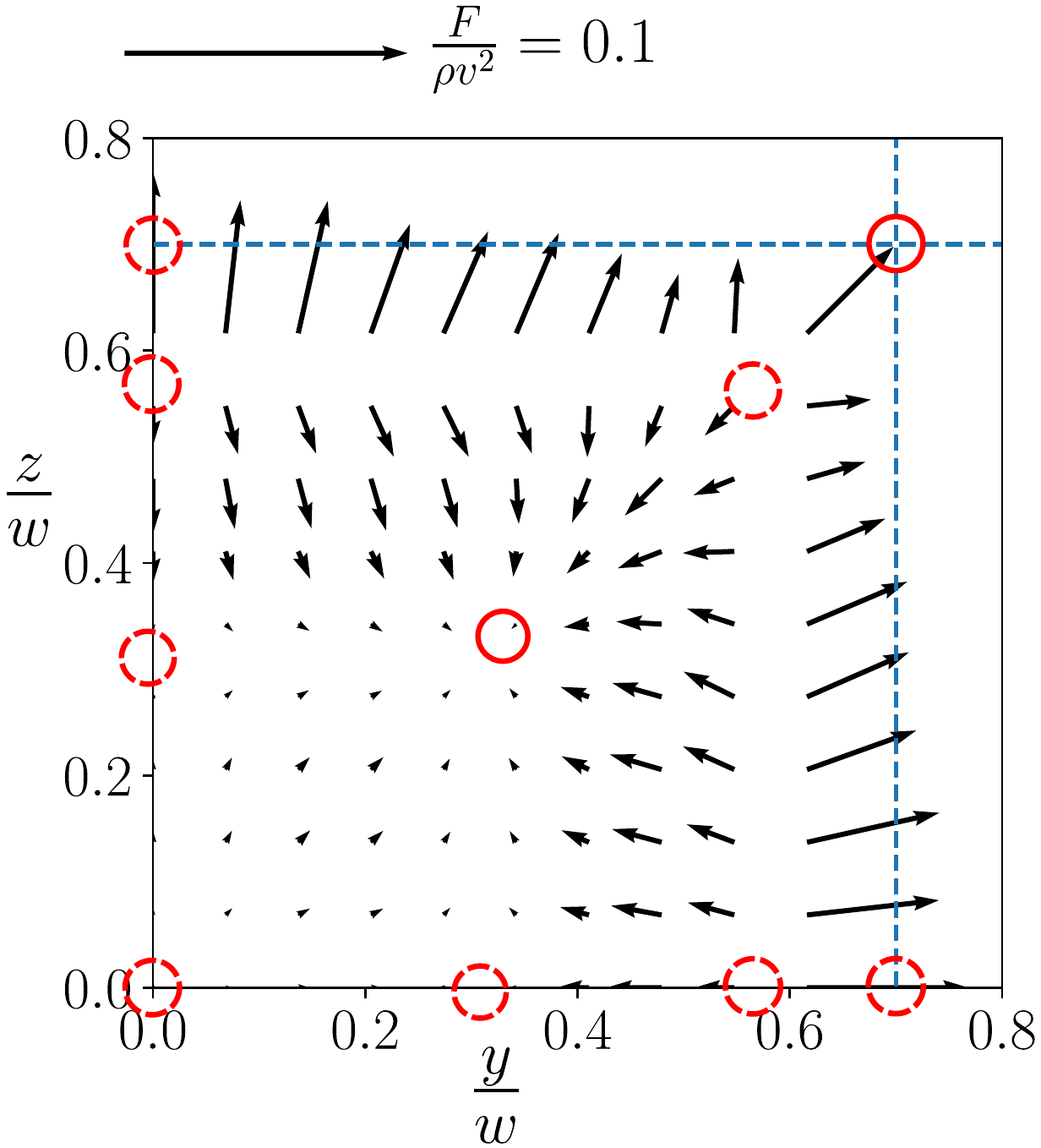}
        \caption{}
    \end{subfigure}%
    ~
     \begin{subfigure}[b]{.5\linewidth}
        \centering
        \includegraphics[height=1.9in]{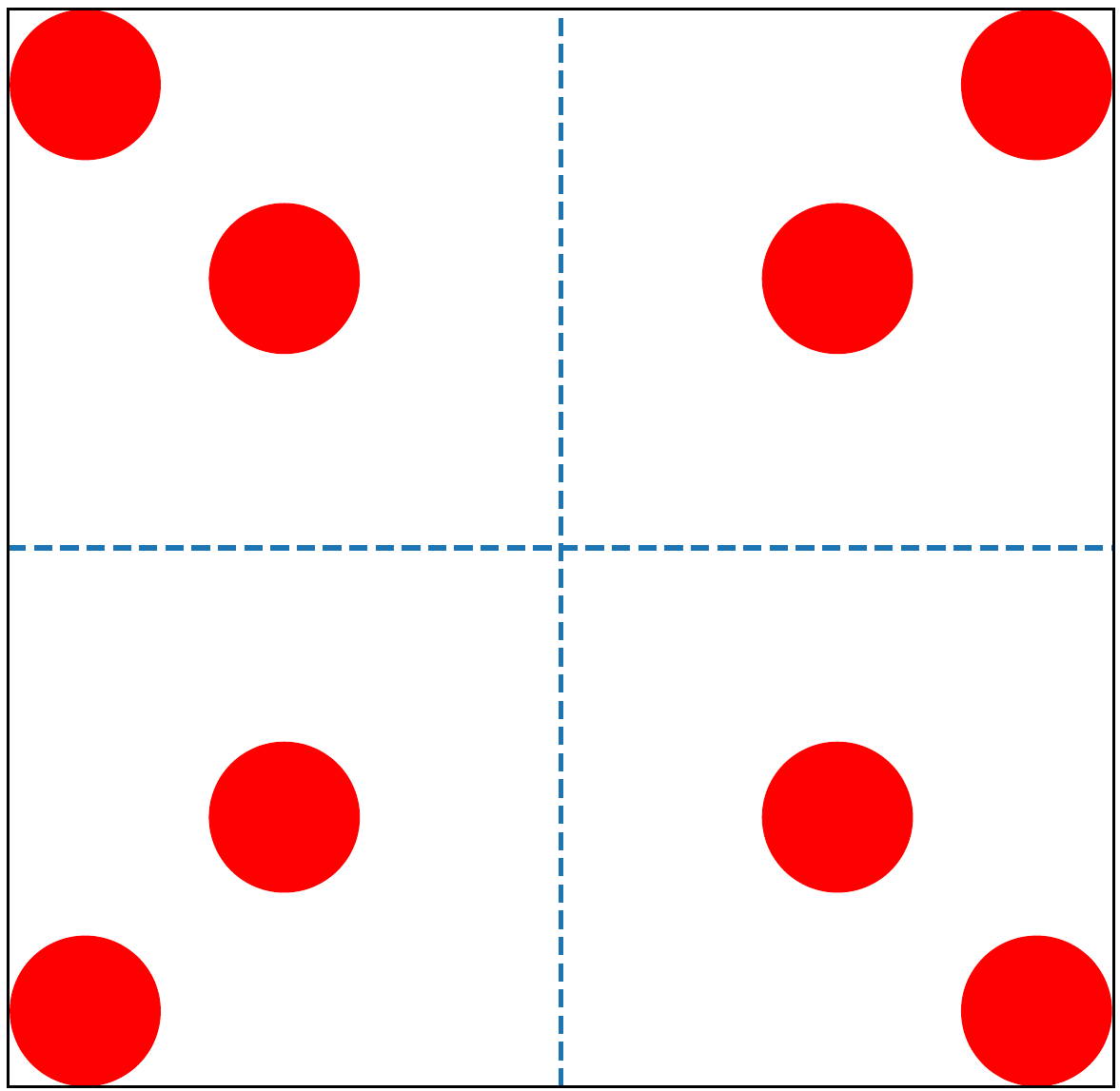}
        \caption{}
    \end{subfigure}%
    
    \begin{subfigure}[t]{.5\linewidth}
        \centering
        \includegraphics[height=2.2in]{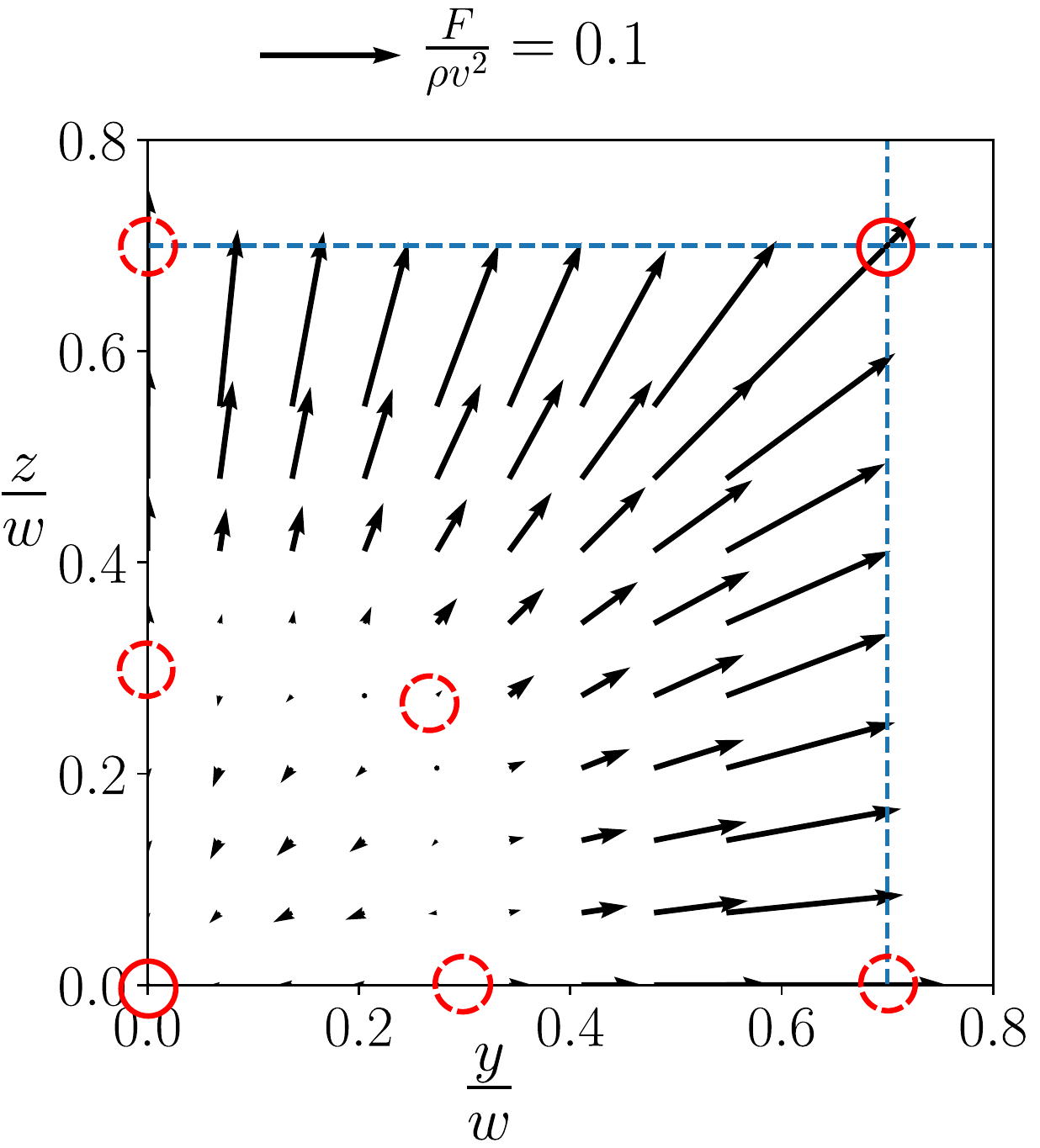}
        \caption{}
    \end{subfigure}%
    ~
    \begin{subfigure}[b]{.5\linewidth}
        \centering
        \includegraphics[height=1.9in]{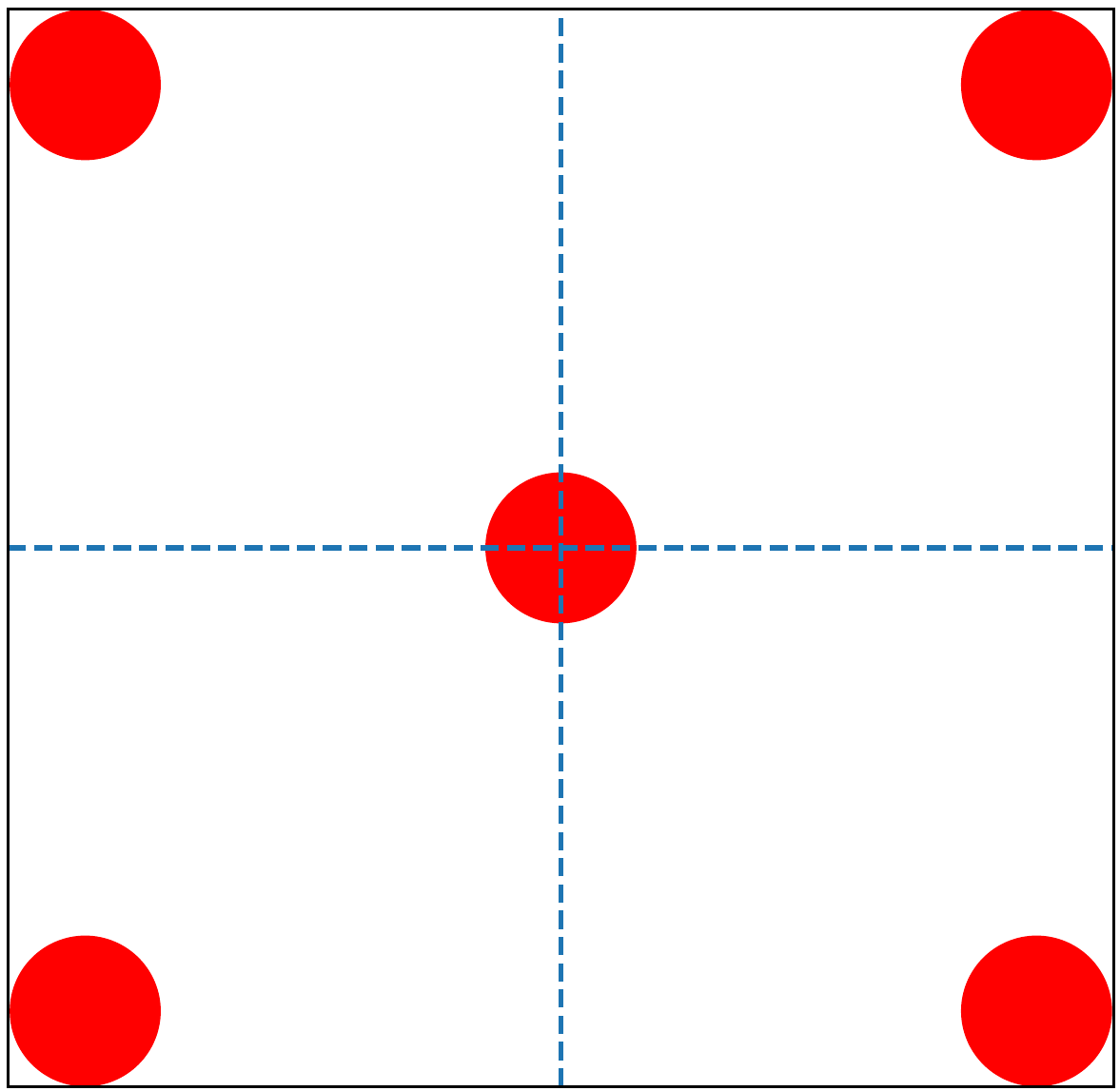}
        \caption{}
    \end{subfigure}%
    \caption{Force-map accross the microchannel for $Re=5$ and  (a) $Wi=0$ and (b) the focusing pattern (stable equilibrium positions) at $Wi=0$, (c) force-map for $Wi=0.1$ and (d) the focusing pattern for $Wi=0.1$, (e) force-map for $Wi=0.5$ and (f) the focusing pattern at $Wi=0.5$.     }
    \label{L_1}
\end{figure*}

\begin{figure*}[ht!]
    \centering
\begin{subfigure}[t]{.5\linewidth}
        \centering
        \includegraphics[height=2.8in]{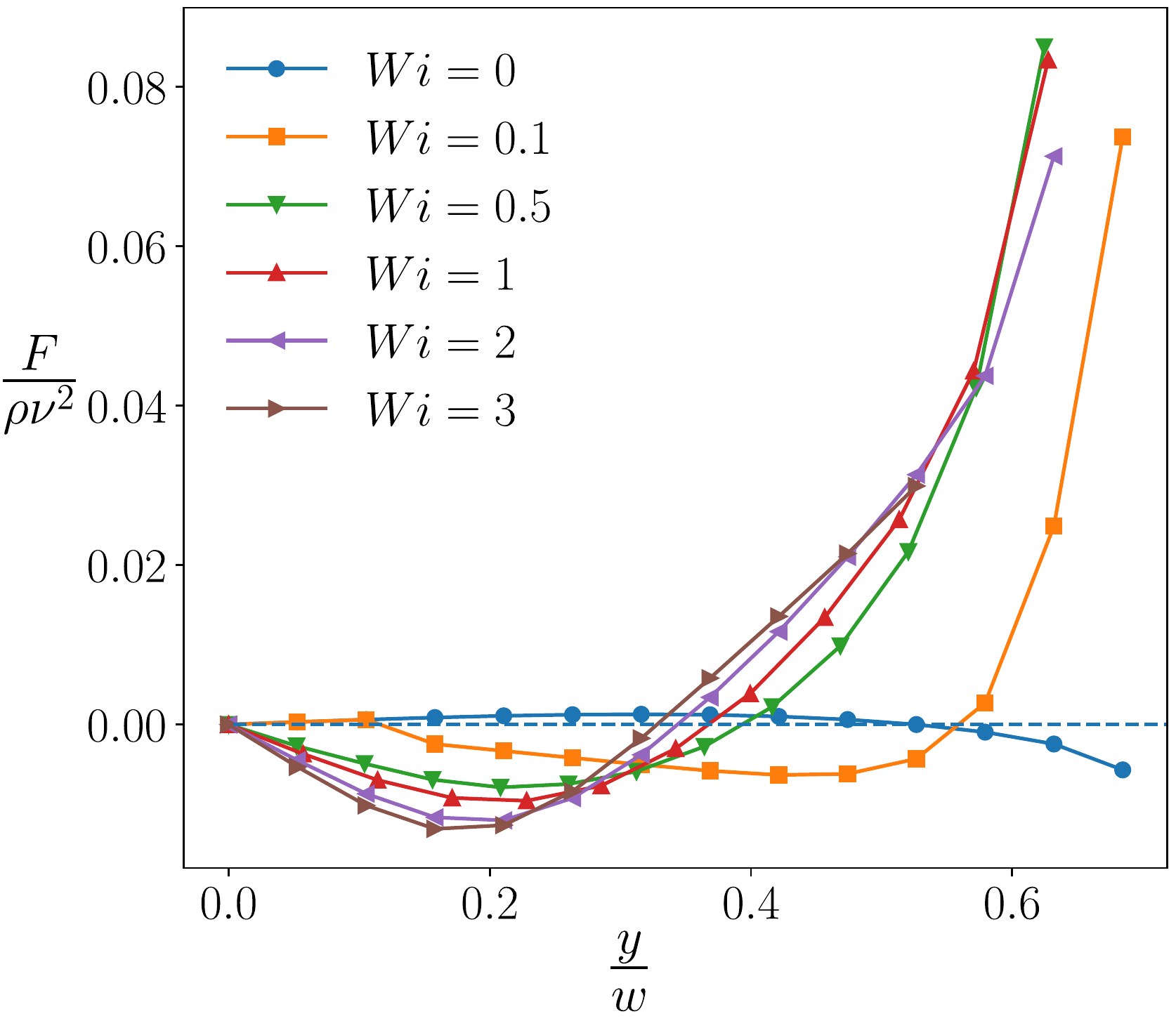}
        \caption{}
    \end{subfigure}%
    ~
    \begin{subfigure}[t]{.5\linewidth}
        \centering
        \includegraphics[height=2.8in]{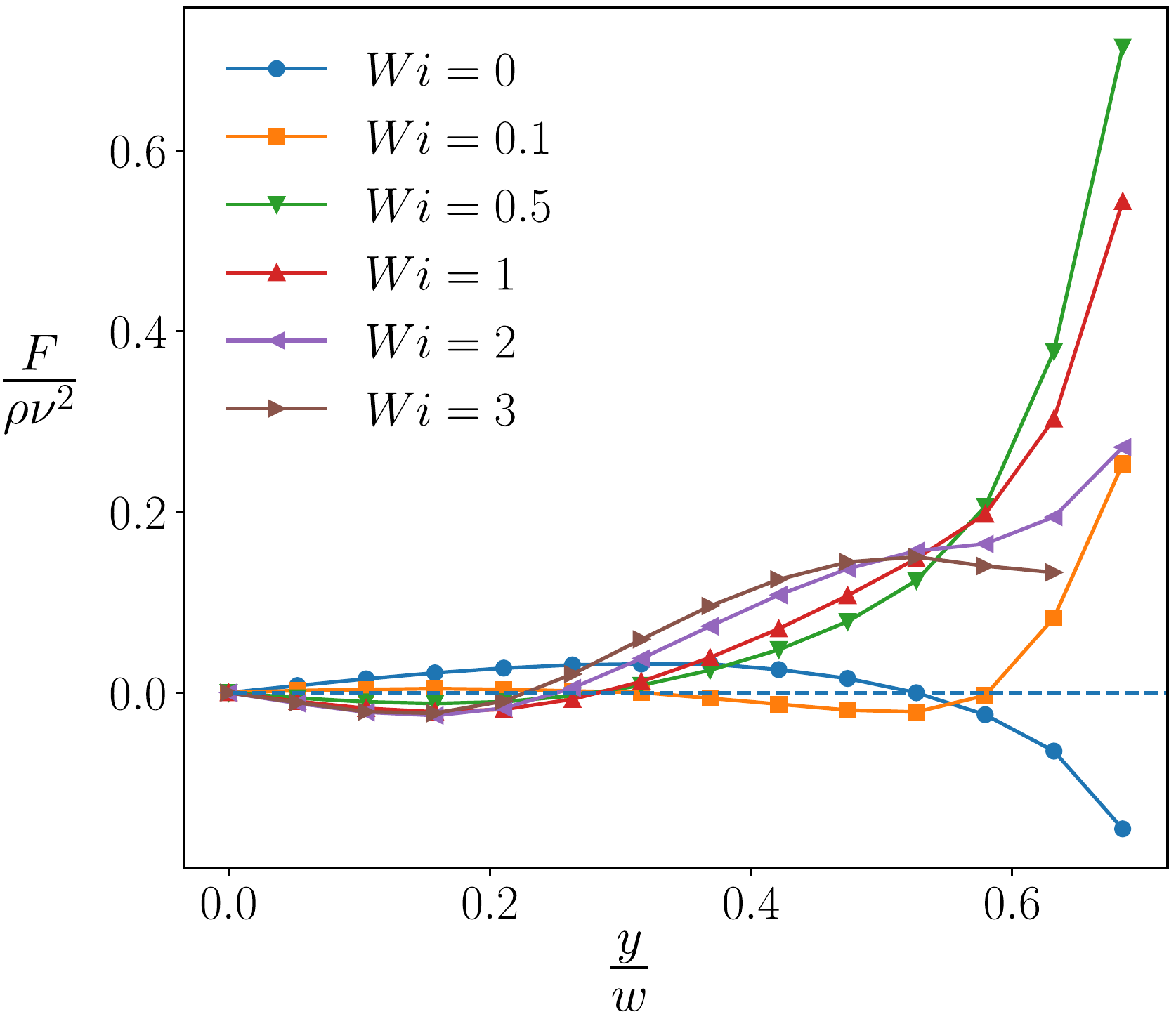}
        \caption{}
    \end{subfigure}
\caption{Lateral force profile along the main axis for (a) $Re=1$ and (b) $Re=5$. }
    \label{L_2}
\end{figure*}

In order to find the location and stability of the equilibrium points in a viscoelastic fluid, the force field experienced by the particle is calculated. The particle size is fixed at $\frac{a}{w}=0.3$ and the Reynolds number is set to $Re=5$ in this section. Due to the symmetry of the problem, the force field is illustrated only for one quarter of the channel cross section. Figure \ref{L_1}(a) shows the lateral force profile acting on the particle at $Re=5$ and $Wi=0$.           
In this figure, the location of the equilibrium points are indicated by red circles. The observed profile explains the particle configuration observed in \cite{choi2011lateral}, \cite{yang2011sheathless} and \cite{di2009particle} As shown in this figure, the radially directed forces drive the particle away from the center and the wall and push the particle toward an annulus ring across the channel marked with the dashed line. Hence, the particle primarily moves in the radial direction to reach the annulus ring. This motion is followed by the migration along the ring to focus at its equilibrium position along the main axes\cite{prohm2014feedback}. Considering the force-map illustrated in Fig. \ref{L_1}(a), there are four stable equilibrium points along the main axes and four unstable equilibrium points along the diagonal of the channel. The results indicate that the lateral force is also zero at the channel center, however, the force field around the channel center implies that this equilibrium point is unstable. The focusing pattern of particles for this force map that can be observed in experiments is presented in Fig. \ref{L_1}(b). In order to further investigate the location and stability of the equilibrium points, the lateral force profile along the y-direction (main axis of the microchannel, i.e., $z=0$ and $y=0$ lines in the cross section) is shown in Fig. \ref{L_2} for $Re=1$ and $5$ and a wide range of $Wi$ number.  
According to Fig. \ref{L_2}(b), the force profile for $Re=5$ and $Wi=0$ crosses the dashed horizontal line at two points (one at the center and the other at $\sim0.52w$). These points are identified as equilibrium points on the main axis and correspond to the locations where the lateral force is zero in Fig. \ref{L_1}(a).  Stability of equilibrium points depends on the slope of the force profile at those locations. Hence, the center of channel is an unstable equilibrium point due to its positive slope, while the off-center equilibrium point is stable due to the negative slope of the force profile in that region. Increasing the $Wi$ number changes the force field and consequently alters the focusing pattern of the particle. Figure \ref{L_1}(c) shows the lateral force acting on the particle for $Re=5$ and $Wi=0.1$. In  the region constrained between two dashed curves  the particle moves toward the center, while in the outer regions the particle moves  toward the walls. A significant change in the force direction can be observed in the near wall region, where it is directed toward the wall in a viscoelastic fluid as opposed to a Newtonian fluid, where the particle is pushed away from the wall. Therefore, the corner becomes a basin of attraction for the particle at $Wi=0.1$. According to the observed force-map, the expected focusing pattern is illustrated in Fig. \ref{L_1}(d). This result also rationalizes the particle behavior found in \cite{seo2014lateral}, where the particles aggregate at the corner and on the diagonal of the channel. The induced lateral force results in the appearance of two equilibrium points along the diagonal line among which the one closer to the channel center is stable and the other one demonstrates an unstable behavior.
Furthermore, there are three unstable equilibrium points on the main axis whose locations are shown in the force profile presented in Fig. \ref{L_2}(b). According to this profile, the channel center and the equilibrium point near the wall are unstable due to the positive slope of the force profile. The middle equilibrium point is also identified as an unstable point despite the negative slope of the curve. This behavior is attributed to the positive lateral force along the z-direction, which pushes the particle away from the main axis and turns this point to a sub-stable equilibrium point.

Figure \ref{L_1}(e) illustrates the force field for $Re=5$ and $Wi=0.5$. In this case the channel can be divided into two regions by a separatix. The region closer to the channel center (indicated by the region inside the dashed curve) attracts the particle toward the centerline, whereas the outer region pushes the particle toward the corner of the microchannel. Hence, only the channel center is stable and other equilibrium points along the diagonal of the channel and the main axes are unstable. It should be noted that the corner is also a basin of attraction for the particles. The corresponding focusing pattern is plotted in Fig. \ref{L_1}(f). The calculated force-map indicates the reason for the particle focusing pattern observed in previous experimental and numerical studies for the cases where the inertial effect is small and the elastic force is dominant\cite{seo2014lateral,del2017edge,yang2011sheathless,villone2013particle}. Figure \ref{L_2}(b) also shows a negative lateral force near the central region and a negative slope for the force profile at the center that leads to stability of the channel center. Another point that should be noted is the change in the force profile near the wall region as the lateral force decreases with increasing the elasticity effect for $Wi$ above $0.5$ in Fig. \ref{L_2}(b). As shown in Fig. \ref{L_2}, the change in $Wi$ number alters the force profile significantly, leading to various focusing patterns for the particles. This effect is more visible for smaller inertial effects as illustrated in Fig. \ref{L_2}(a) ($Re=1$).  Increasing the $Wi$ number changes the convexity of the force profile along the main axis. In the Newtonian case ($Wi=0$), the convexity of the force is negative along the entire axis, while it becomes positive for higher $Wi$ numbers. Furthermore, the channel center which is unstable for $Wi=0$, becomes stable for any $Wi$ number above $0.5$ and the corner becomes the basin of attraction. This change is observed for both $Re=1$ and $5$. According to Fig. \ref{L_2}(a) and (b), the location at which the force profile crosses the horizontal line in the range of $Wi>0.5$ shifts toward the channel center with increasing elasticity. Hence, the size of the separatix shrinks with $Wi$. This behavior can explain the results found in previous studies\cite{del2017edge,villone2013particle,seo2014lateral} as larger fraction of particles get attracted to the corner with increasing the elastic  effects ($Wi$).      
\begin{figure*}[h!]
    \centering
\begin{subfigure}[t]{.5\linewidth}
        \centering
        \includegraphics[height=2.8in]{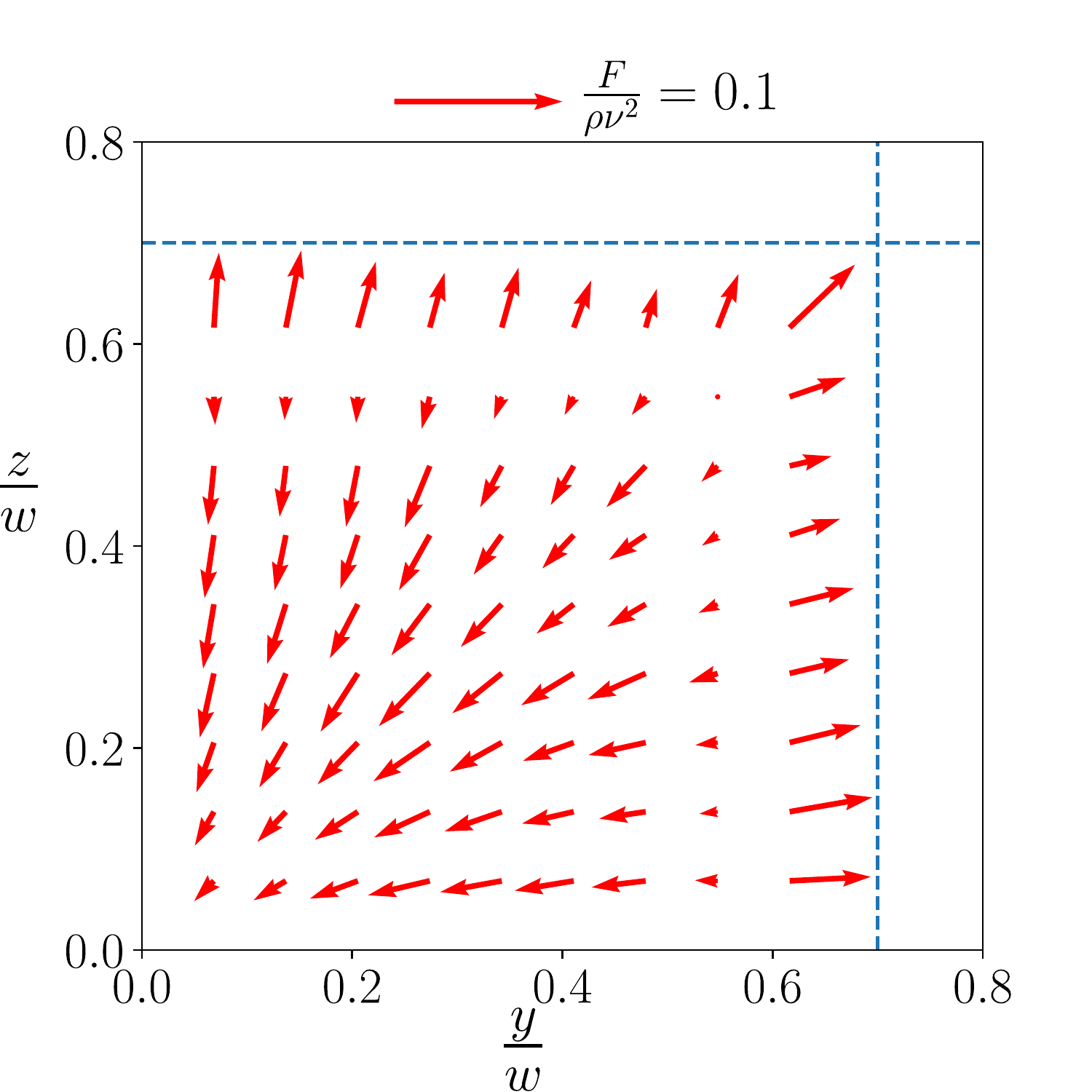}
        \caption{}
    \end{subfigure}%
    ~
    \begin{subfigure}[t]{.5\linewidth}
        \centering
        \includegraphics[height=2.8in]{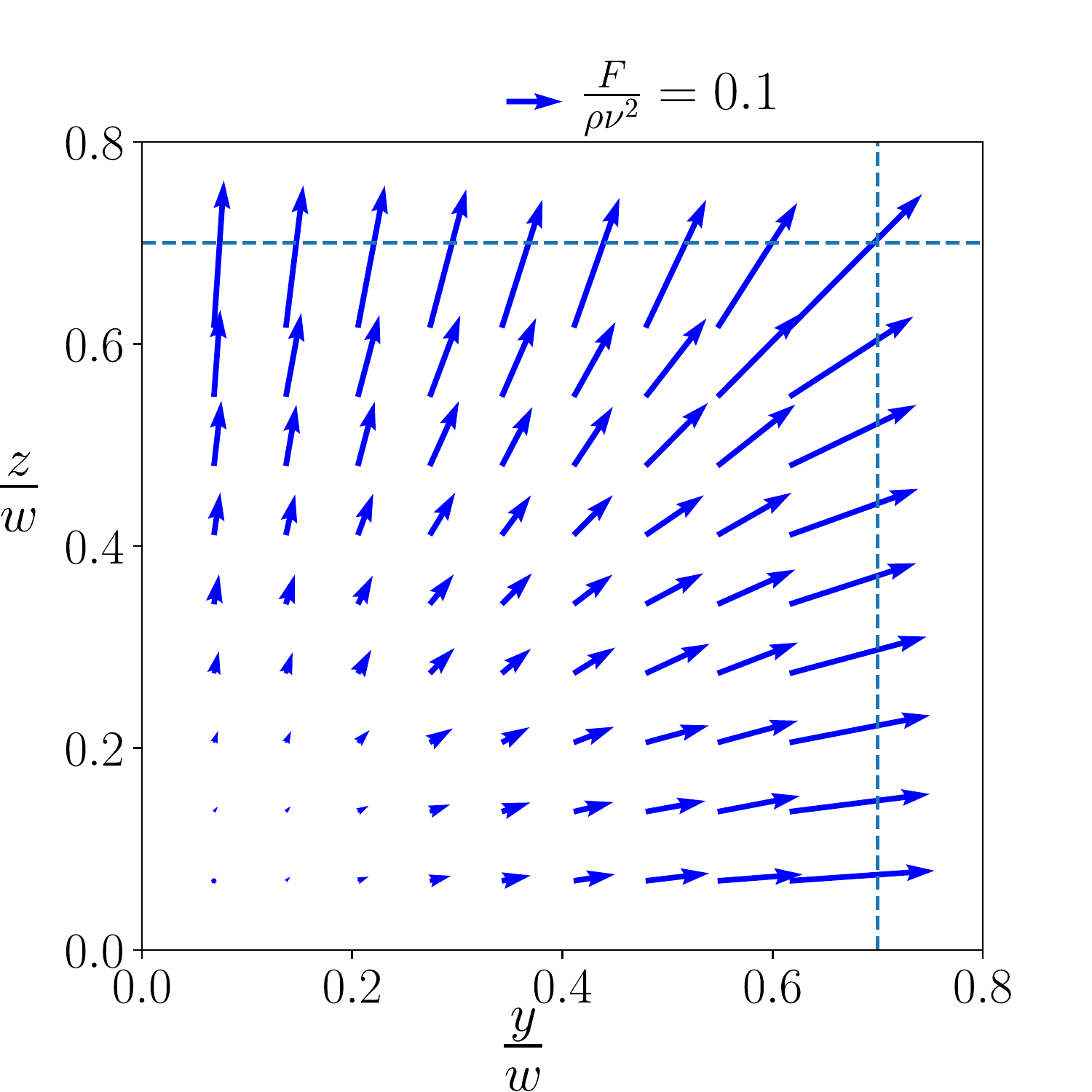}
        \caption{}
    \end{subfigure}
\caption{Distribution of (a) elastic force and (b) inertial force at $Re=5$ and $Wi=0.5$ }
    \label{L_3}
\end{figure*}

In order to investigate the effect of elasticity, particularly in the case where inertia is small, we split the lateral force plotted in Fig. \ref{L_1}(e) into two components: (i) elastic force ($F_{el}$) and (ii) inertial force ($F_{in}$). Figure \ref{L_3} shows the distribution of elastic and inertial forces for $Re=5$ and $Wi=0.5$. The elastic force (represented in Fig. \ref{L_3}(a)) drives the particle toward the center in the entire  channel unless the particle is positioned near the wall region, where the elastic force direction reverses and the particle is pushed toward the wall. Contrary to the elastic force profile, the inertial force shown in Fig. \ref{L_3}(b) repels the particle from the center across the entire channel, however, the magnitude of the inertial force becomes negligible compared to that of elastic force for in the region close to the channel center. Hence, we observed a trapping region formed near the center of the microchannel and the particles that fall outside this region migrate to the corner. The comparison between elastic and inertial forces shown in Fig. \ref{L_3} can explain the observed experimental results in previous studies. It should be noted that the magnitude of the elastic and inertial forces are significantly different  for some cases and we use different force scale bars, which are shown in the figures.  
\begin{figure*}[h!]
    \centering
\begin{subfigure}[t]{.5\linewidth}
        \centering
        \includegraphics[height=2.8in]{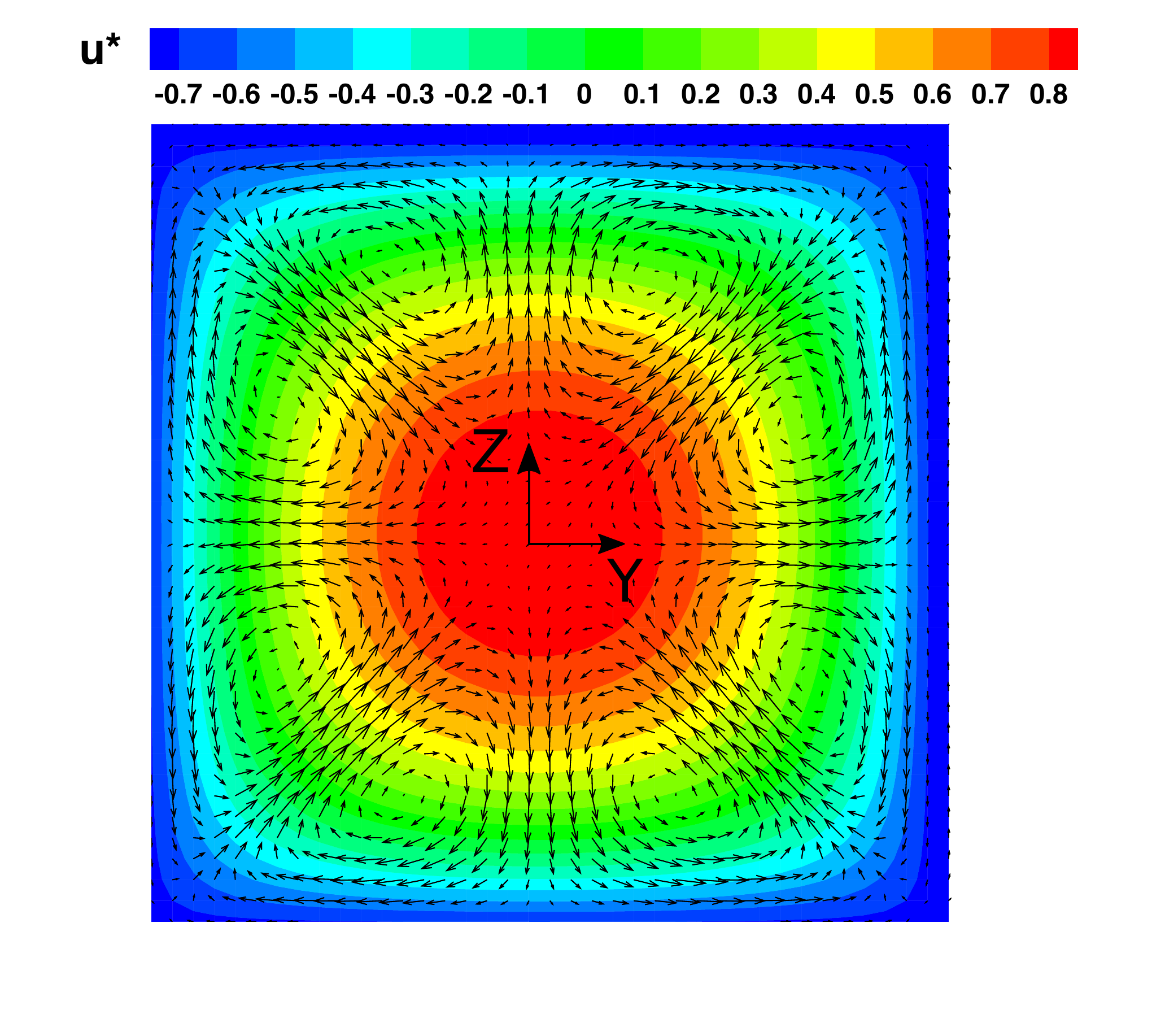}
        \caption{}
    \end{subfigure}%
    ~
    \begin{subfigure}[t]{.5\linewidth}
        \centering
        \includegraphics[height=2.8in]{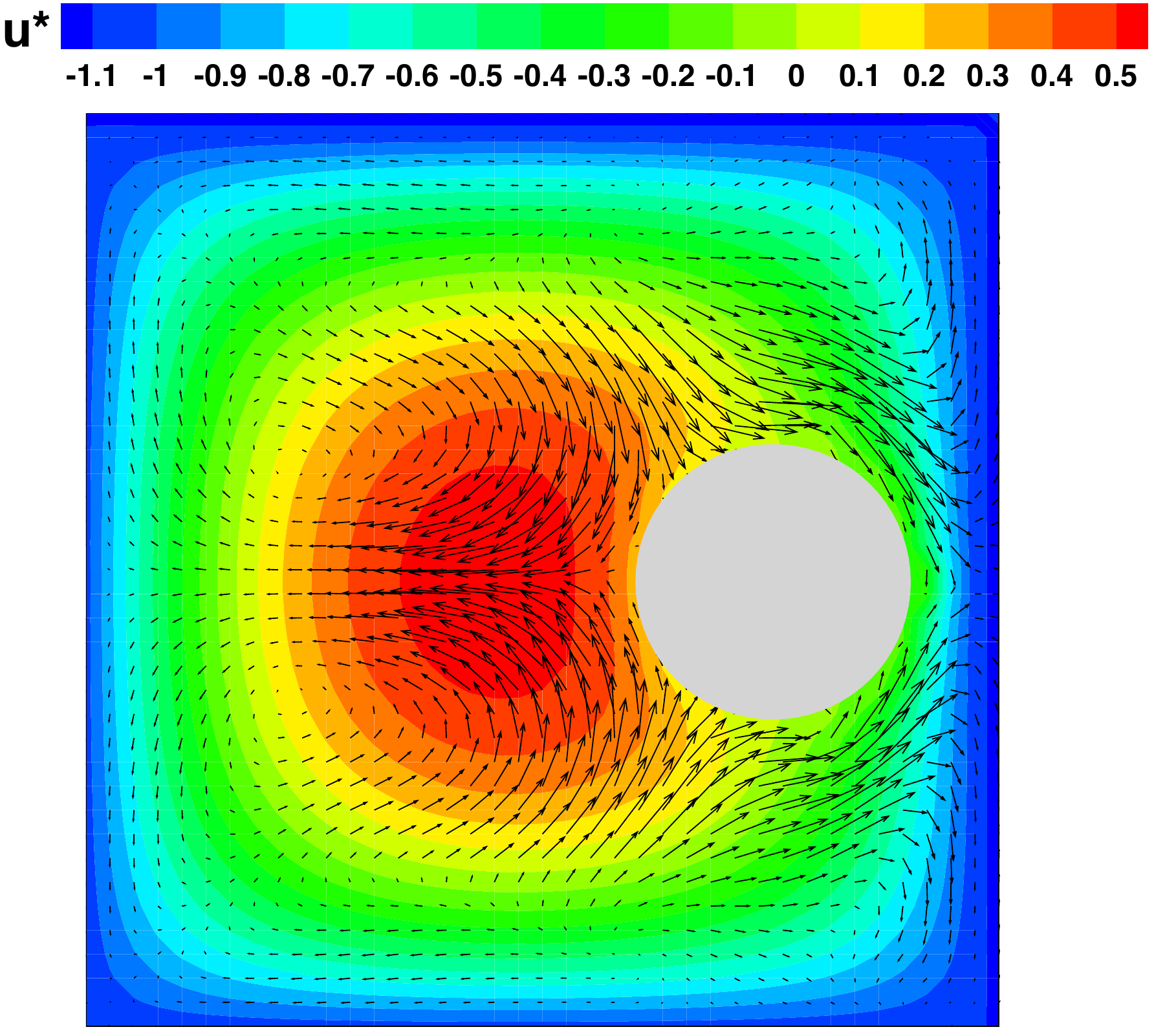}
        \caption{}
    \end{subfigure}
\caption{Velocity profile at (a) the channel inlet and (b) the location of particle center at $Re=5$ and $Wi=3$. The stream-wise velocity is defined as $u^*=\frac{u}{U_0}$.} 
    \label{L_4}
\end{figure*}
Figure \ref{L_2} shows that increasing $Wi$ number increases the magnitude of the lateral force near the channel center (for $Wi>0.1$).  This effect accelerates the transverse migration of the particle across the microchannel and leads to a shorter critical length for the microchannel required for complete focusing of the suspending particles\cite{d2017particle,villone2013particle}. The second normal stress is non-zero in a Giesekus fluid modeled in this paper and it causes a secondary flow  across the microchannel which affects the particle migration in the channel\cite{li2015dynamics,lu2017particle,villone2013particle}.  In order to show the effect of the secondary flow, the velocity profile is illustrated in Figure \ref{L_4} at the channel inlet (far from the particle) and the location of particle center at $Re=5$ and $Wi=3$. 

\subsection{Migration in an intermediate inertial regime}
\begin{figure*}[h!]
    \centering
    \begin{subfigure}[t]{.5\linewidth}
        \centering
        \includegraphics[height=2.3in]{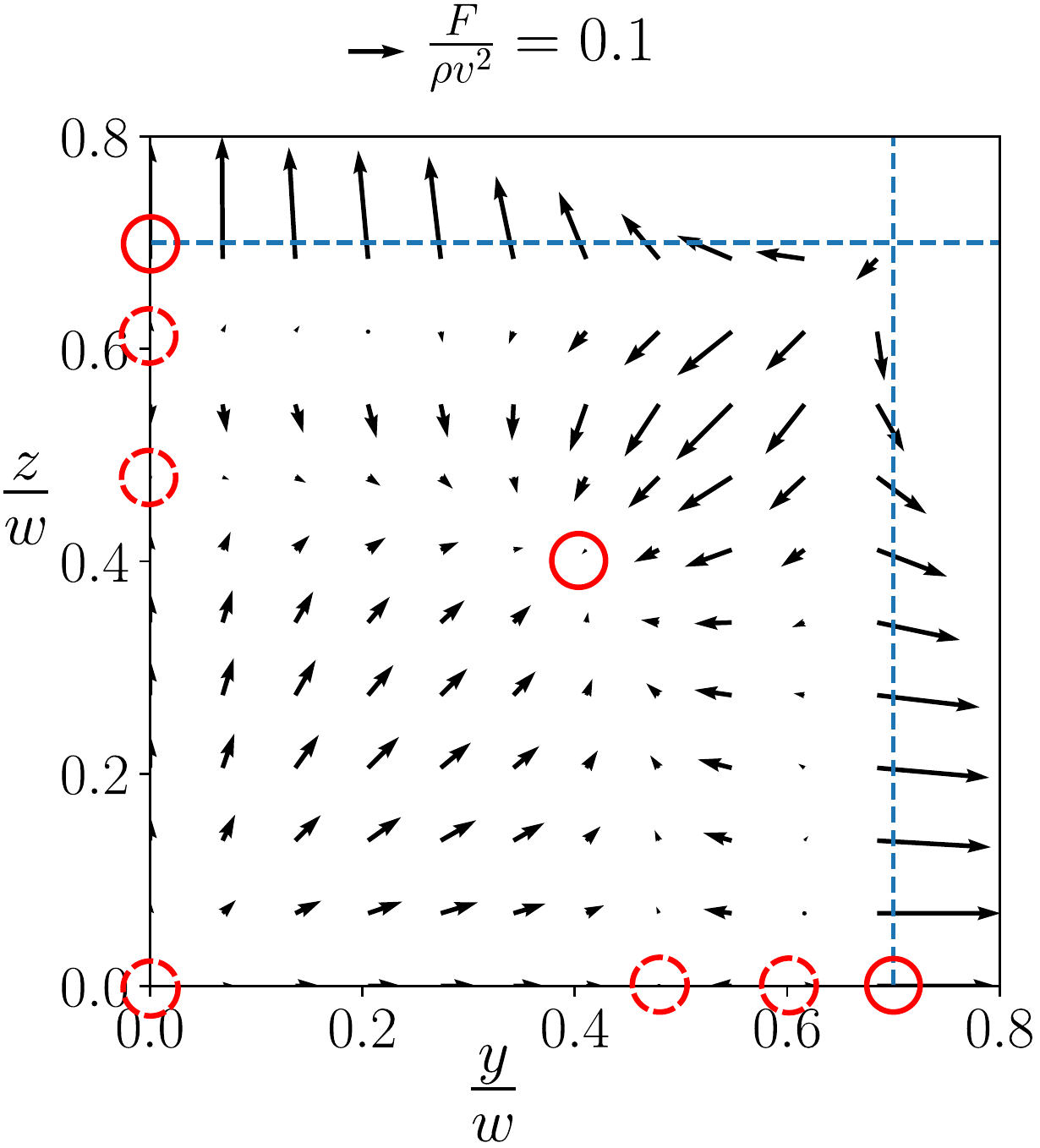}
        \caption{}
    \end{subfigure}%
    ~
    \begin{subfigure}[b]{.5\linewidth}
        \centering
        \includegraphics[height=1.9in]{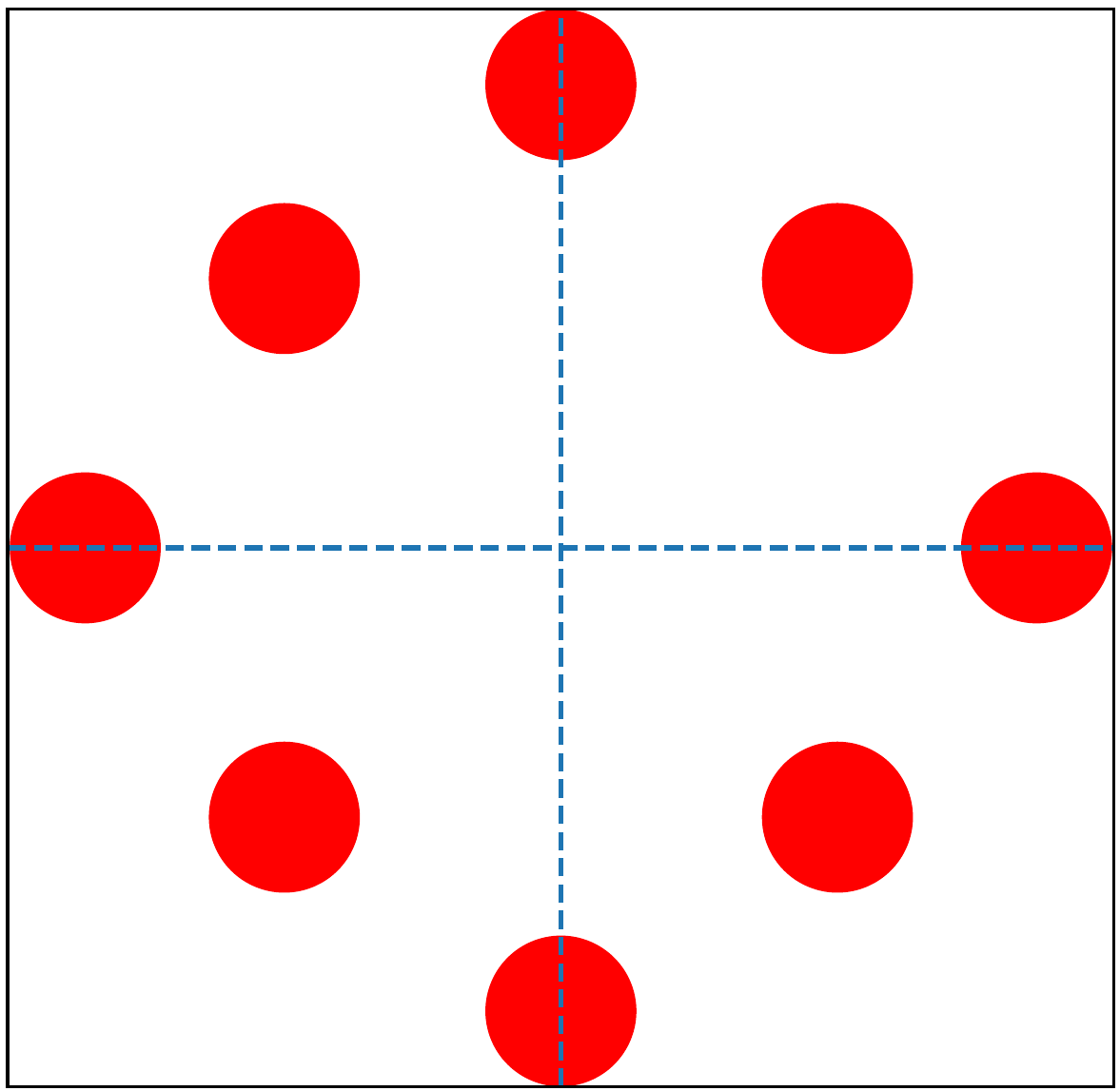}
        \caption{}
    \end{subfigure}
    
     \begin{subfigure}[t]{.5\linewidth}
        \centering
        \includegraphics[height=2.3in]{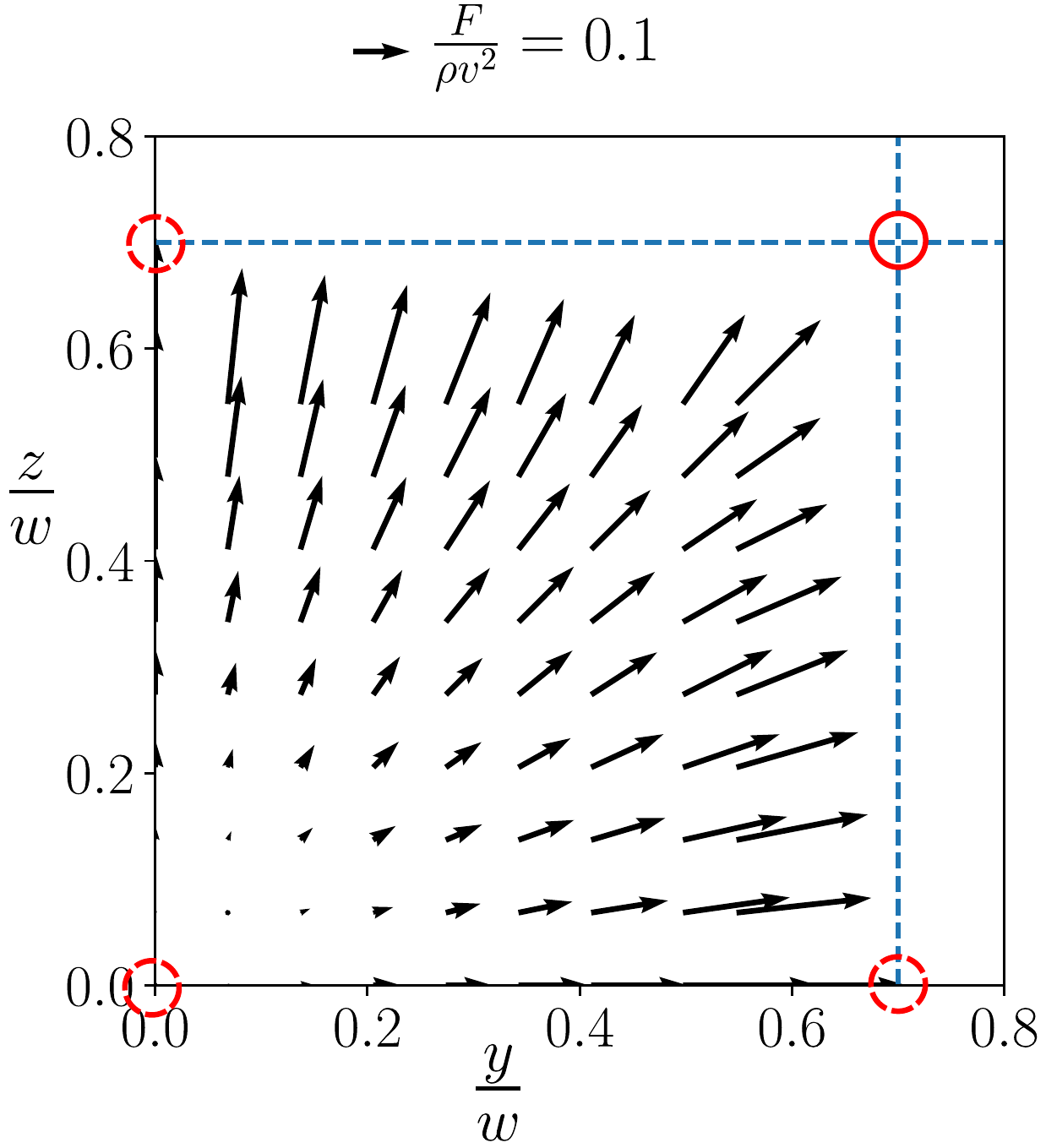}
        \caption{}
    \end{subfigure}%
    ~
     \begin{subfigure}[b]{.5\linewidth}
        \centering
        \includegraphics[height=1.9in]{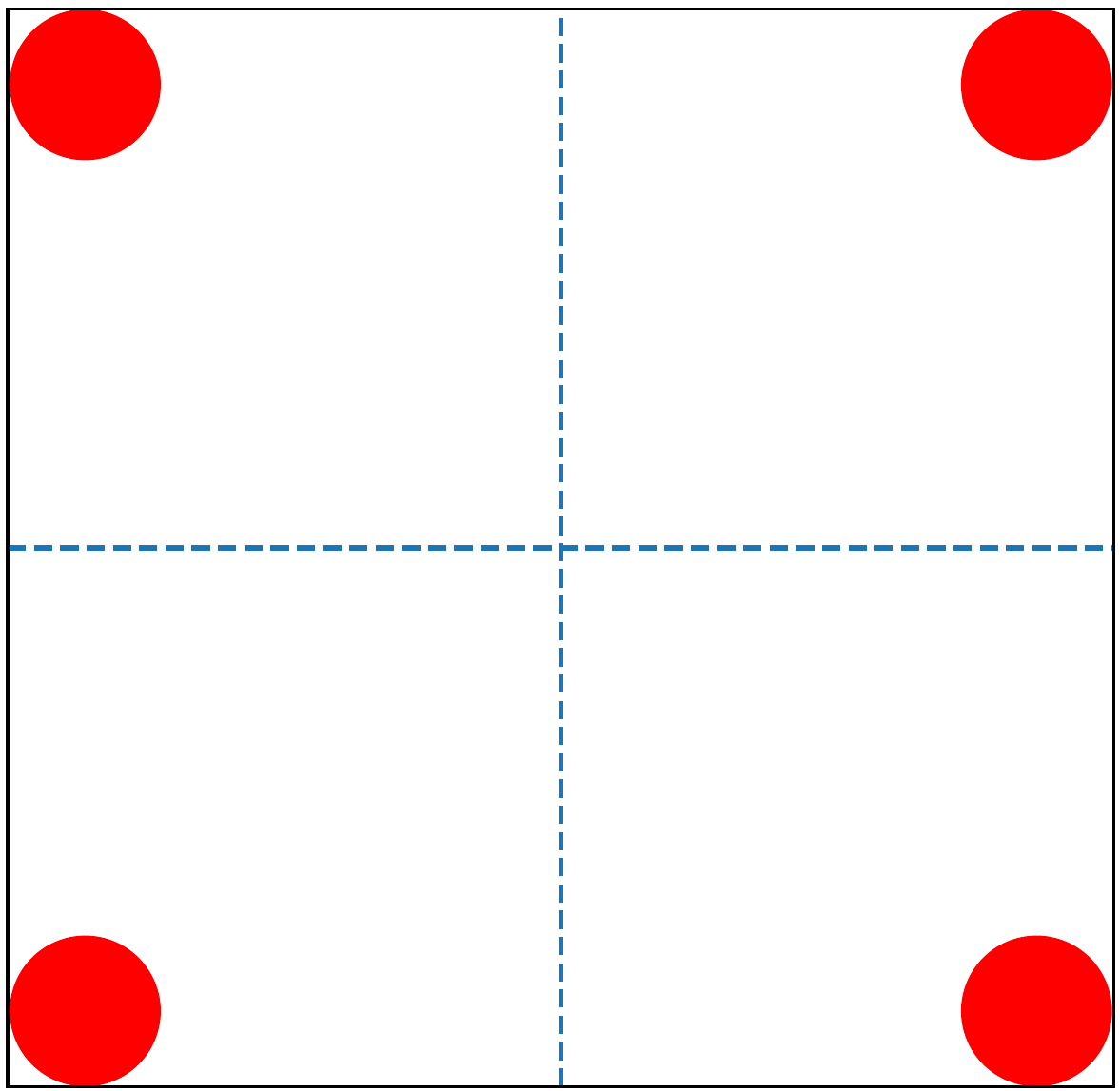}
        \caption{}
    \end{subfigure}%
    
    \begin{subfigure}[t]{.5\linewidth}
        \centering
        \includegraphics[height=2.3in]{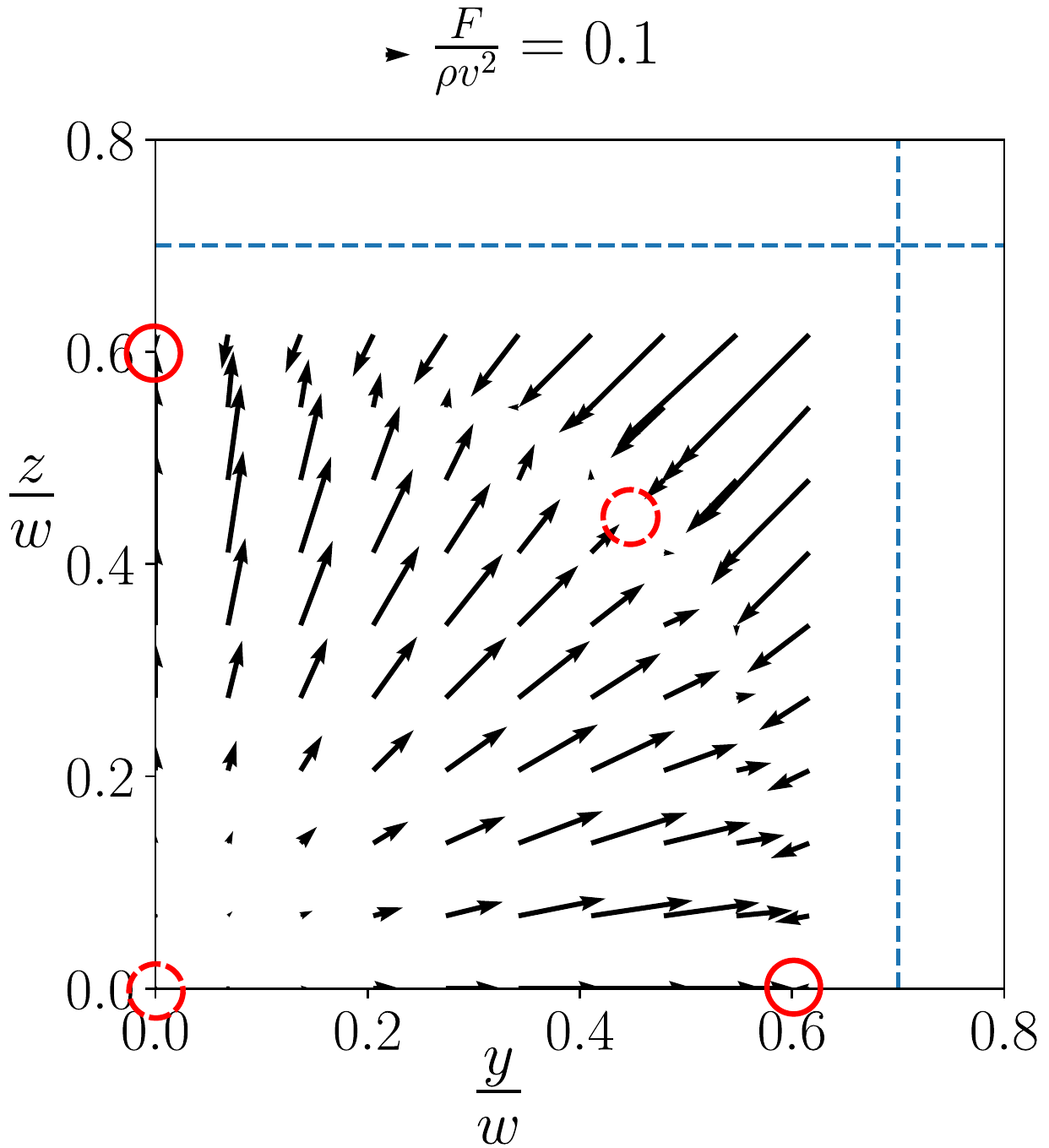}
        \caption{}
    \end{subfigure}%
    ~
    \begin{subfigure}[b]{.5\linewidth}
        \centering
        \includegraphics[height=1.9in]{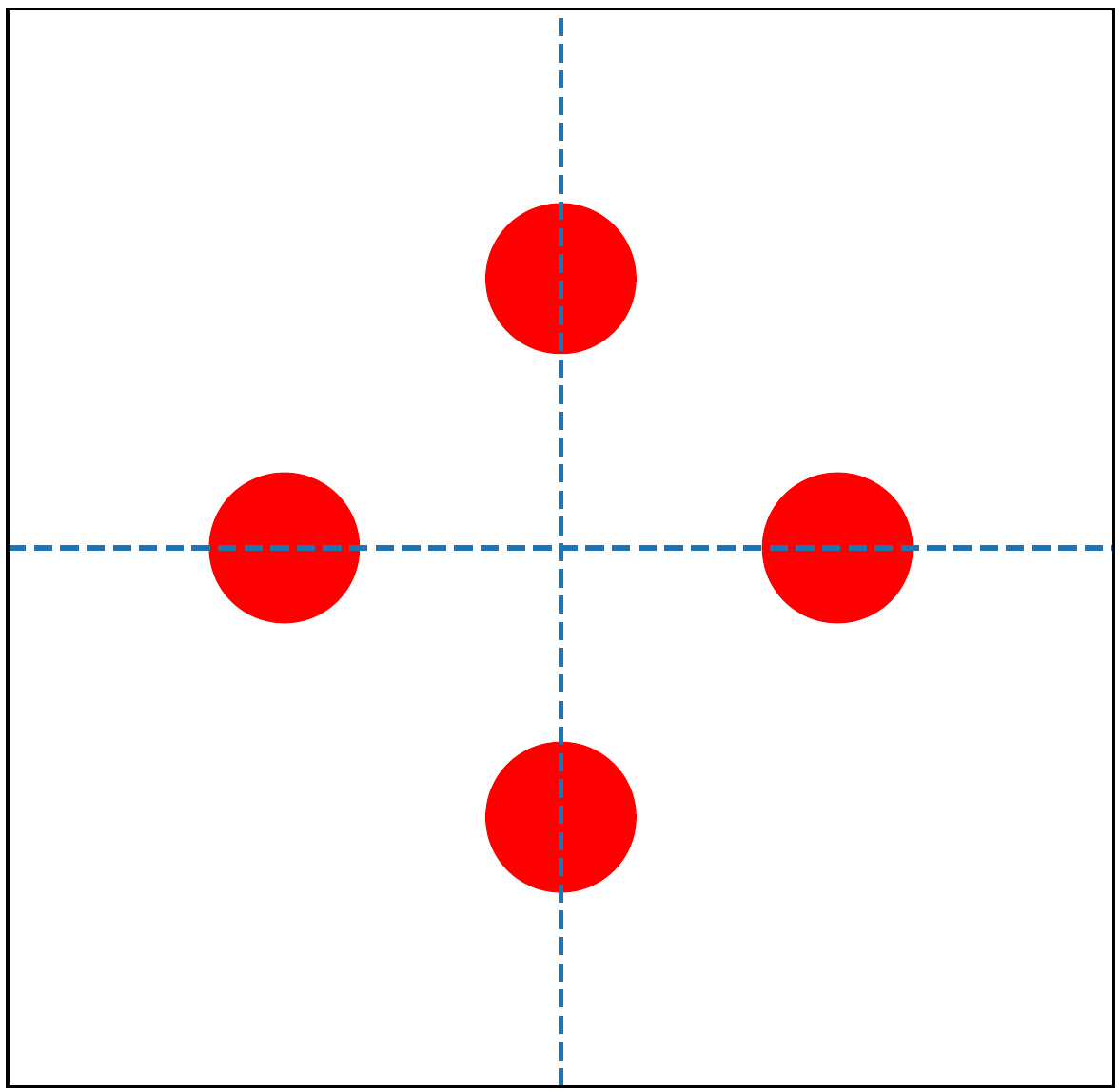}
        \caption{}
    \end{subfigure}%
    \caption{Force-map accross the microchannel for $Re=10$ and  (a) $Wi=0.1$ and (b) the focusing pattern at $Wi=0.1$, (c) force-map for $Wi=0.5$ and (d) the focusing pattern for $Wi=0.5$, (e) force-map for $Wi=3$ and (f) the focusing pattern at $Wi=3$.     }
    \label{M_1}
\end{figure*}

\begin{figure}[h!]
    \centering
        \includegraphics[height=2.8in]{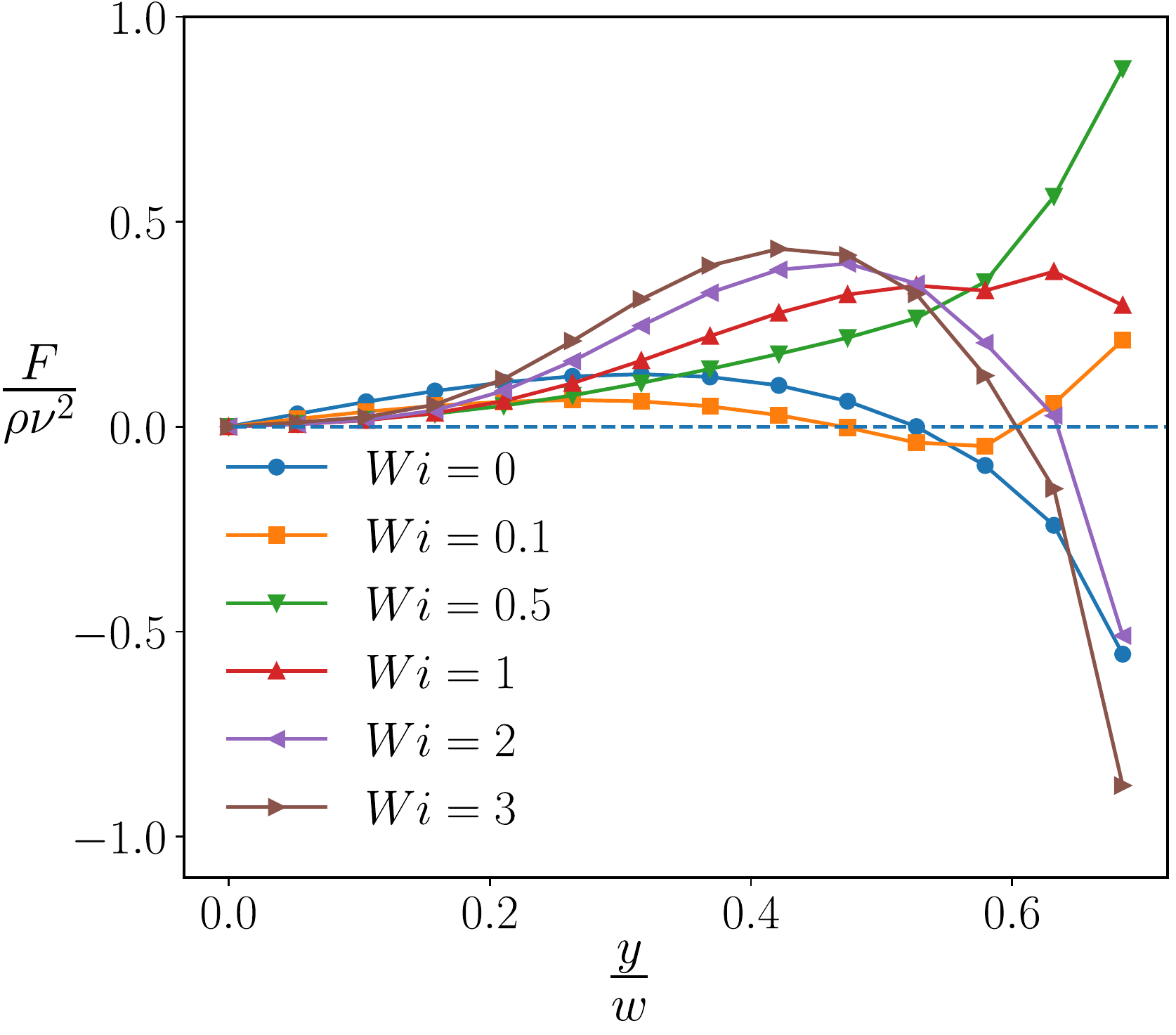}
\caption{Lateral force profile on the main axis for $Re=10$ }
    \label{M_2}
\end{figure}

In this section, we investigate the particle focusing pattern at $Re=10$ for various Weissenberg numbers. The larger inertial effect  leads to a significant change in the generated flow field compared to that of the previous section. This change can be reflected in the induced force-map shown in Fig. \ref{M_1}. Figure \ref{M_1}(a) illustrates the force field at $Wi=0.1$. The noticeable difference between this case and the Newtonian fluid is the induced lateral force near the channel face center that attracts the particle toward the wall leading to the existence of a basin of attraction at the channel face.
Furthermore, the generated lateral force turns the equilibrium point on the diagonal of the channel stable and those on the main axes and channel center  unstable. This phenomenon can be observed in Fig. \ref{M_2}, where the force profile along the main axis is plotted, showing the locations where the lateral force becomes zero and their corresponding stability status. Hence, the expected focal pattern looks like the one illustrated in Fig. \ref{M_1}(b) which has not been discovered in previous studies. Increasing the $Wi$ number changes the force field significantly as displayed in Fig. \ref{M_1}(c). The radially directed lateral force drives the particle toward the channel corner across the entire channel. Hence, the corner becomes the only basin of attraction resulting in the focal pattern presented in Fig. \ref{M_1}(d). Correspondingly, the positive value of the force along the entire main axis for $Wi=0.5$ shown in Fig. \ref{M_2} indicates that the particle is pushed away from the center regardless of its location in the microchannel. The calculated force field shown in Fig. \ref{M_1}(c) explains the particle configuration reported in previous experimental studies\cite{del2017edge,seo2014lateral}. For $Wi=3$, the lateral force has a distribution similar to that of a Newtonian fluid. In this case (shown in Fig. \ref{M_1}(e)) the radially directed force pushes the particle away from the wall and channel center and creates an annulus ring which is similar to the curve marked in Fig. \ref{L_1}(a). Consequently, the predicted focal pattern of the particle shown in Fig. \ref{M_1}(f) is similar to that of a Newtonian fluid.

\begin{figure*}[h!]
    \centering
\begin{subfigure}[t]{0.3\textwidth}
        \centering
        \includegraphics[height=1.83in,keepaspectratio]{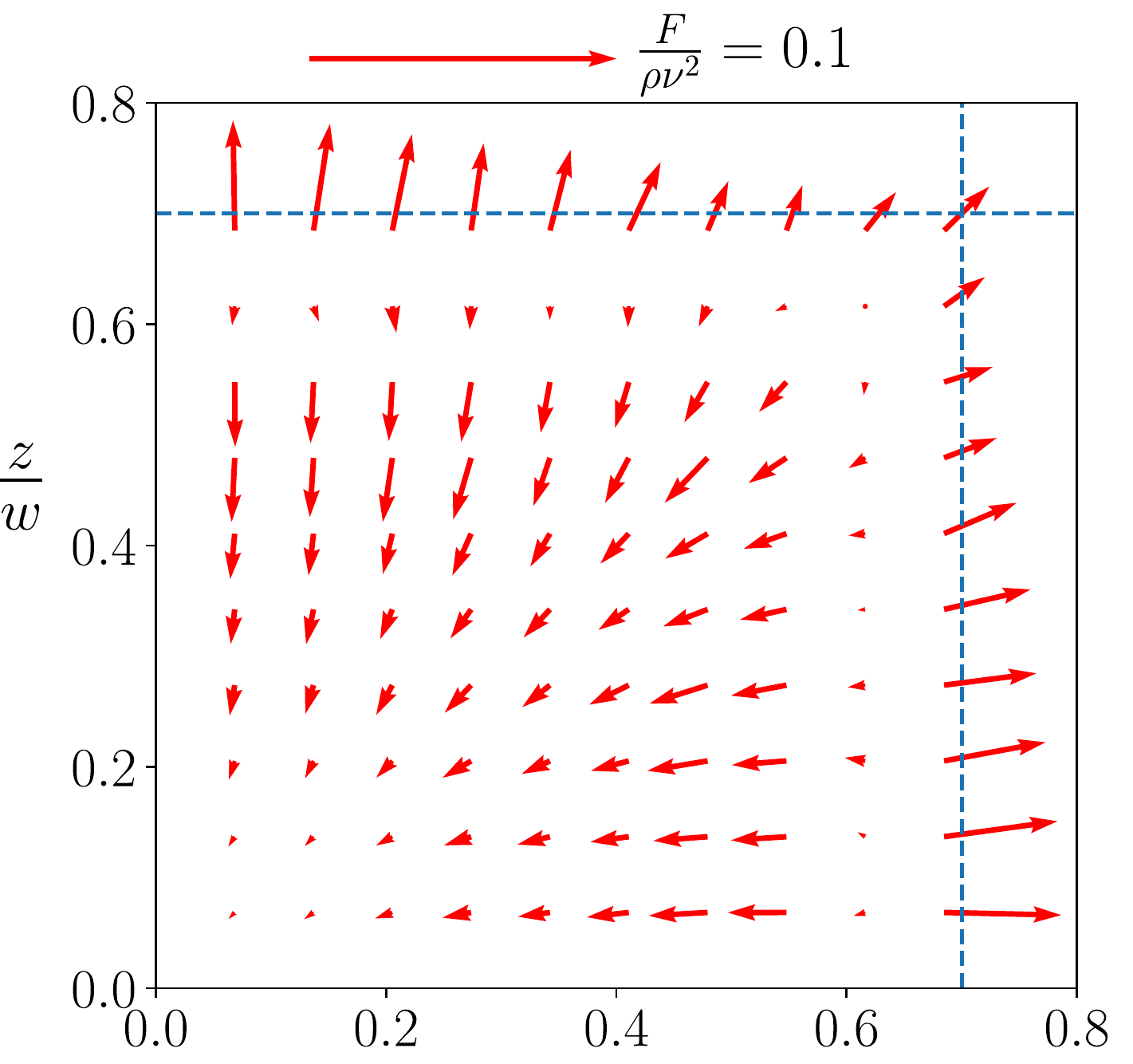}
        \caption{}
    \end{subfigure}%
    ~
    \begin{subfigure}[t]{.35\textwidth}
        \centering
        \includegraphics[height=1.79in,keepaspectratio]{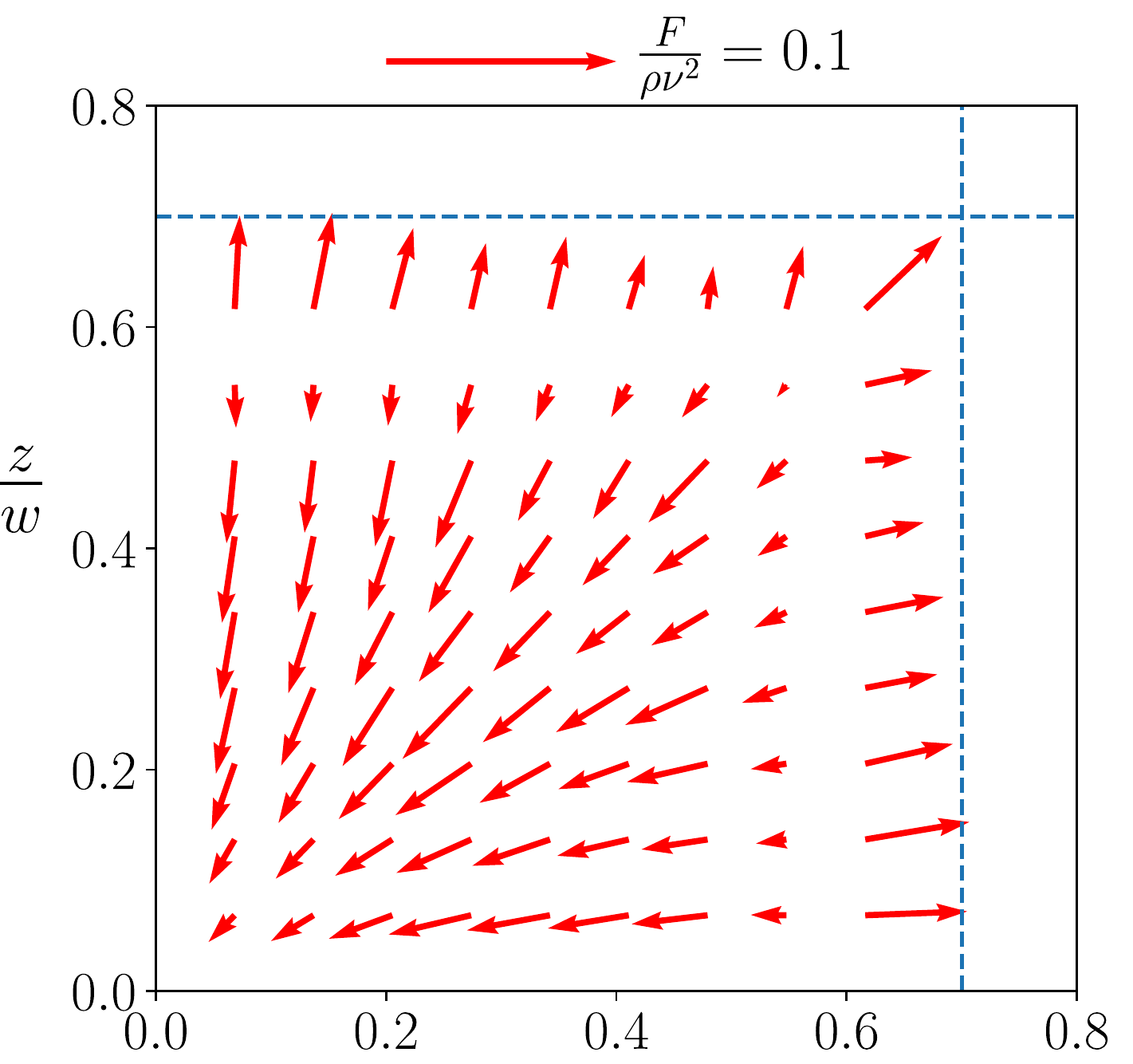}
        \caption{}
    \end{subfigure}
    ~
    \begin{subfigure}[t]{.2\textwidth}
        \centering
        \includegraphics[height=1.79in,keepaspectratio]{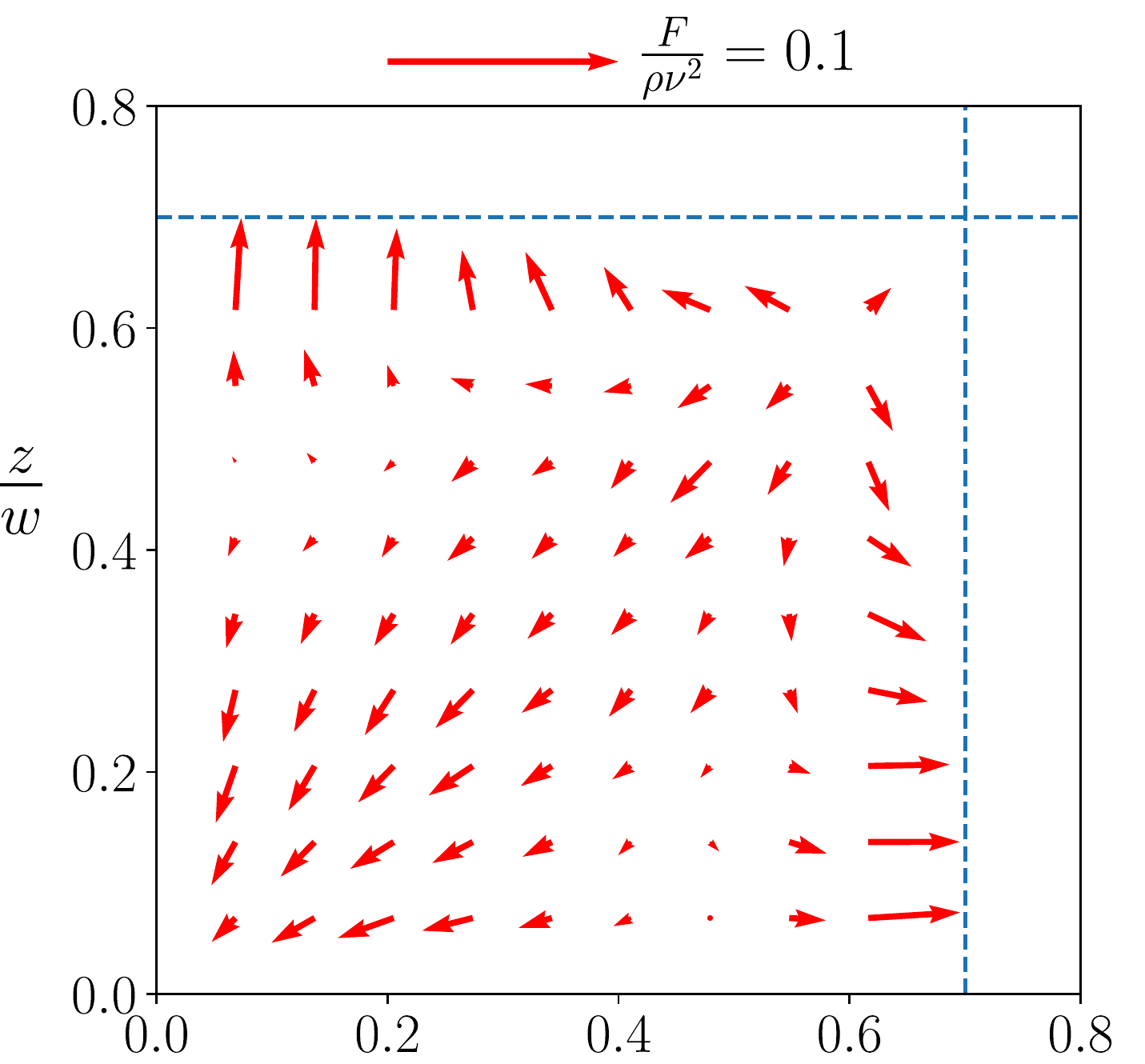}
        \caption{}
    \end{subfigure}
    ~
    \begin{subfigure}[t]{0.3\textwidth}
        \centering
        \includegraphics[height=2.15in,keepaspectratio]{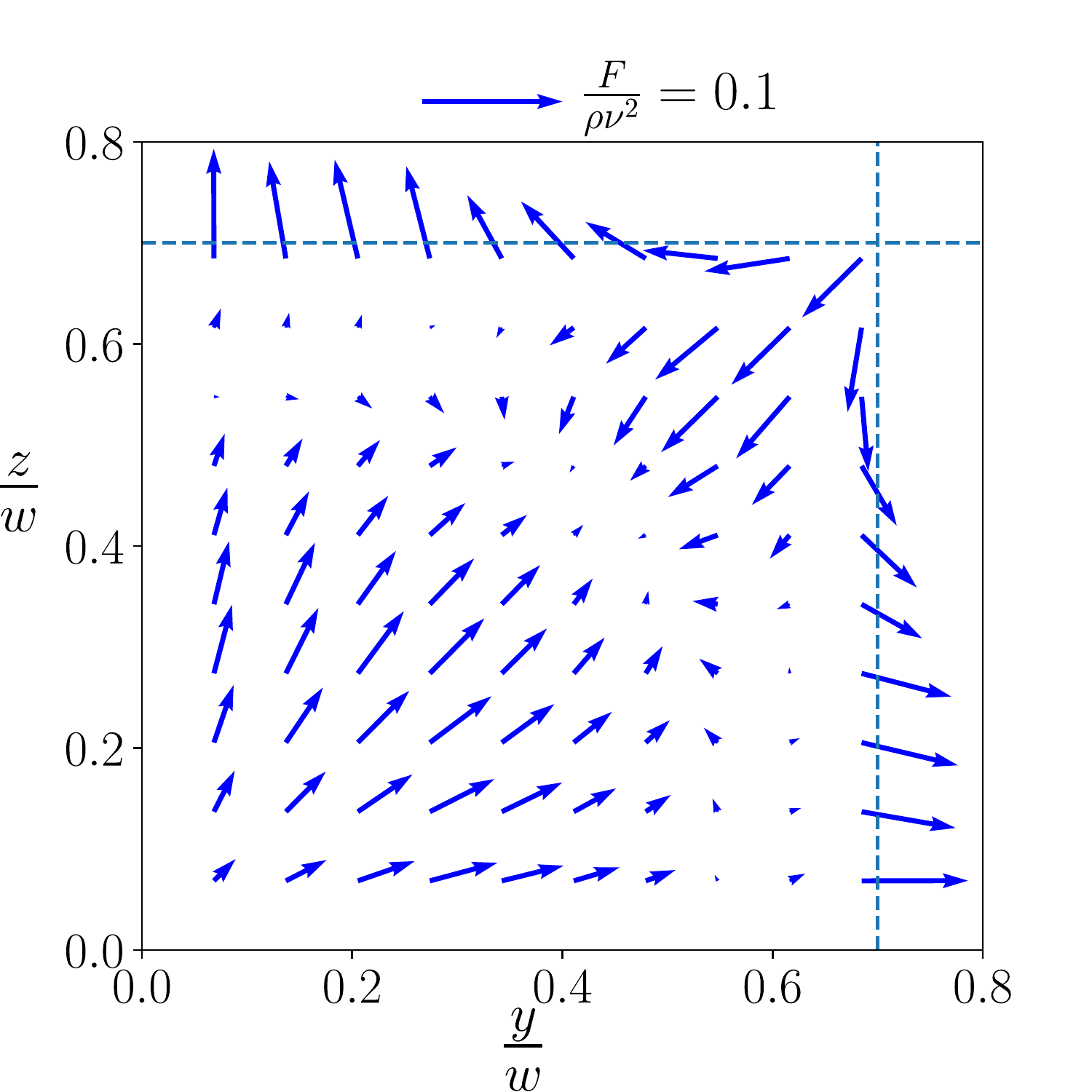}
        \caption{}
    \end{subfigure}%
    ~
    \begin{subfigure}[t]{.35\textwidth}
        \centering
        \includegraphics[height=2.15in]{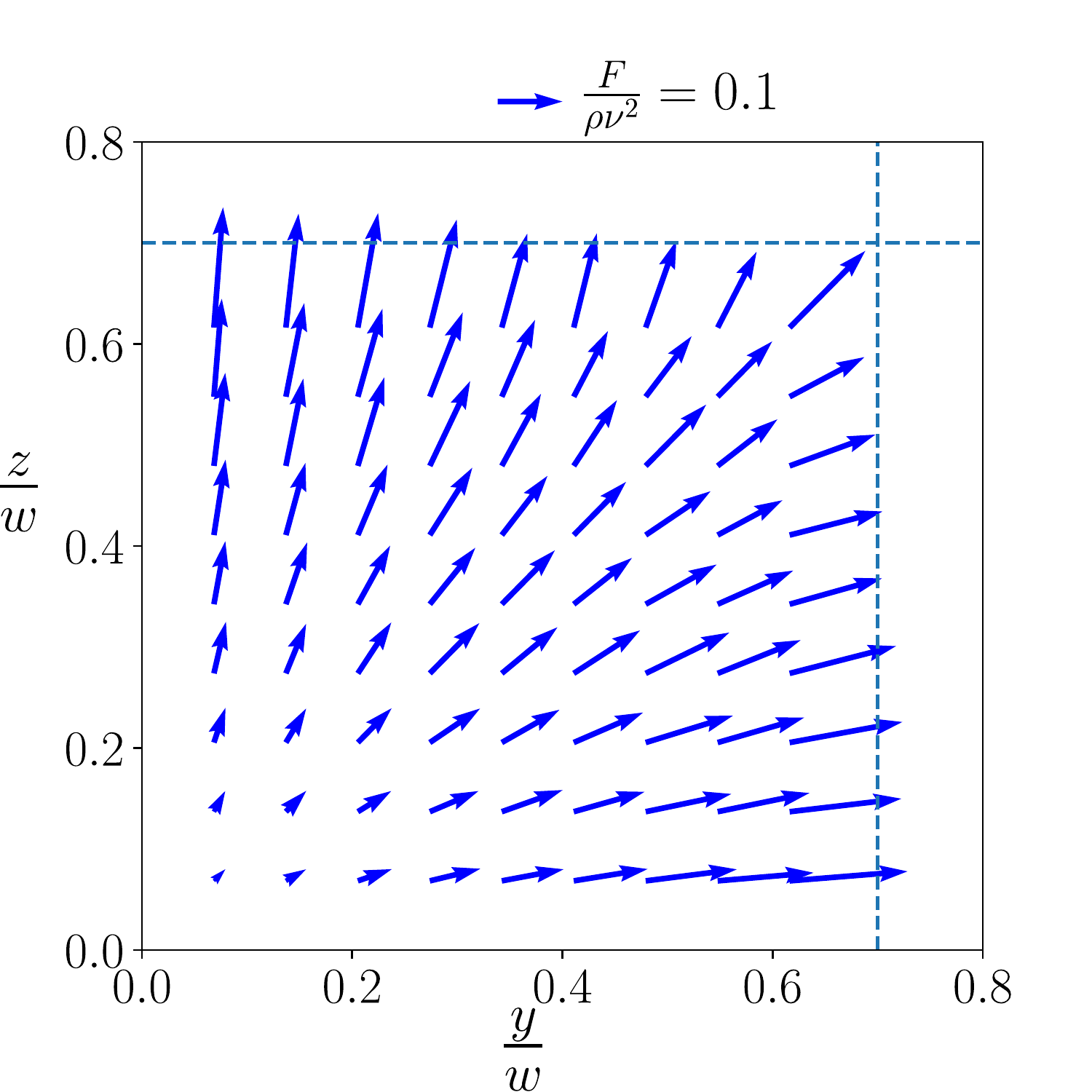}
        \caption{}
    \end{subfigure}
    ~
    \begin{subfigure}[t]{.2\textwidth}
        \centering
        \includegraphics[height=2.15in]{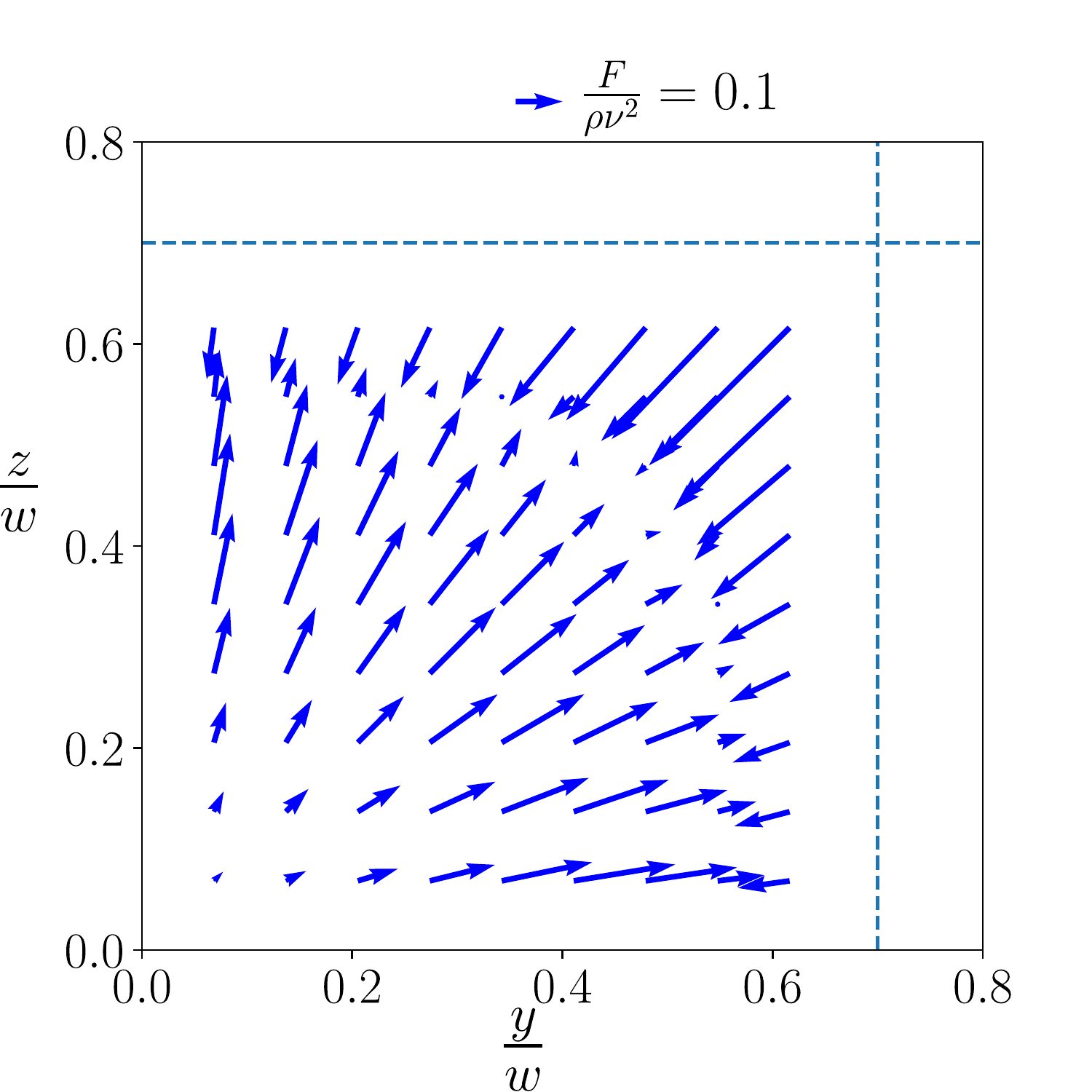}
        \caption{}
    \end{subfigure}
\caption{Distribution of elastic force at $Re=10$ and (a) $Wi=0.1$ (b) $Wi=0.5$ (c) $Wi=3$ and inertial force at (d) $Wi=0.1$ (e) $Wi=0.5$ (f) $Wi=3$  }
    \label{M_3}
\end{figure*}

According to Fig. \ref{M_2}, the fluid elasticity  significantly affects the magnitude of the lateral force such that the force profile near the wall region markedly changes and its direction reverses for high $Wi$ numbers, leading to a stable equilibrium point on the main axes. On the other hand, the slope of the force profile is positive at the center for the entire range of Weissenberg number  explored in this study, implying that the channel center is not a stable position for the particle in this range of parameters. In order to investigate the effect of elasticity on the total lateral force the force-map of inertial and elastic forces are plotted for $Wi=0.1$ to $3$ in Fig. \ref{M_3}.     
The results show that the inertial force dominates the elastic force. The direction of total force matches that of the inertial force, while the elastic force has a similar profile across the entire channel for various values of $Wi$ number. This behavior indicates that changing the elasticity of the fluid changes the velocity field in the channel. Consequently, the inertial force is affected and the resulting focal pattern alters, while the elastic force itself remains relatively unchanged. In other word, the fluid elasticity affects the particle dynamic indirectly by changing the flow field and not directly by means of elastic force.

\subsection{Migration in a high inertial regime}
\begin{figure*}[h!]
    \centering
\begin{subfigure}[t]{.5\linewidth}
        \centering
        \includegraphics[height=2.8in]{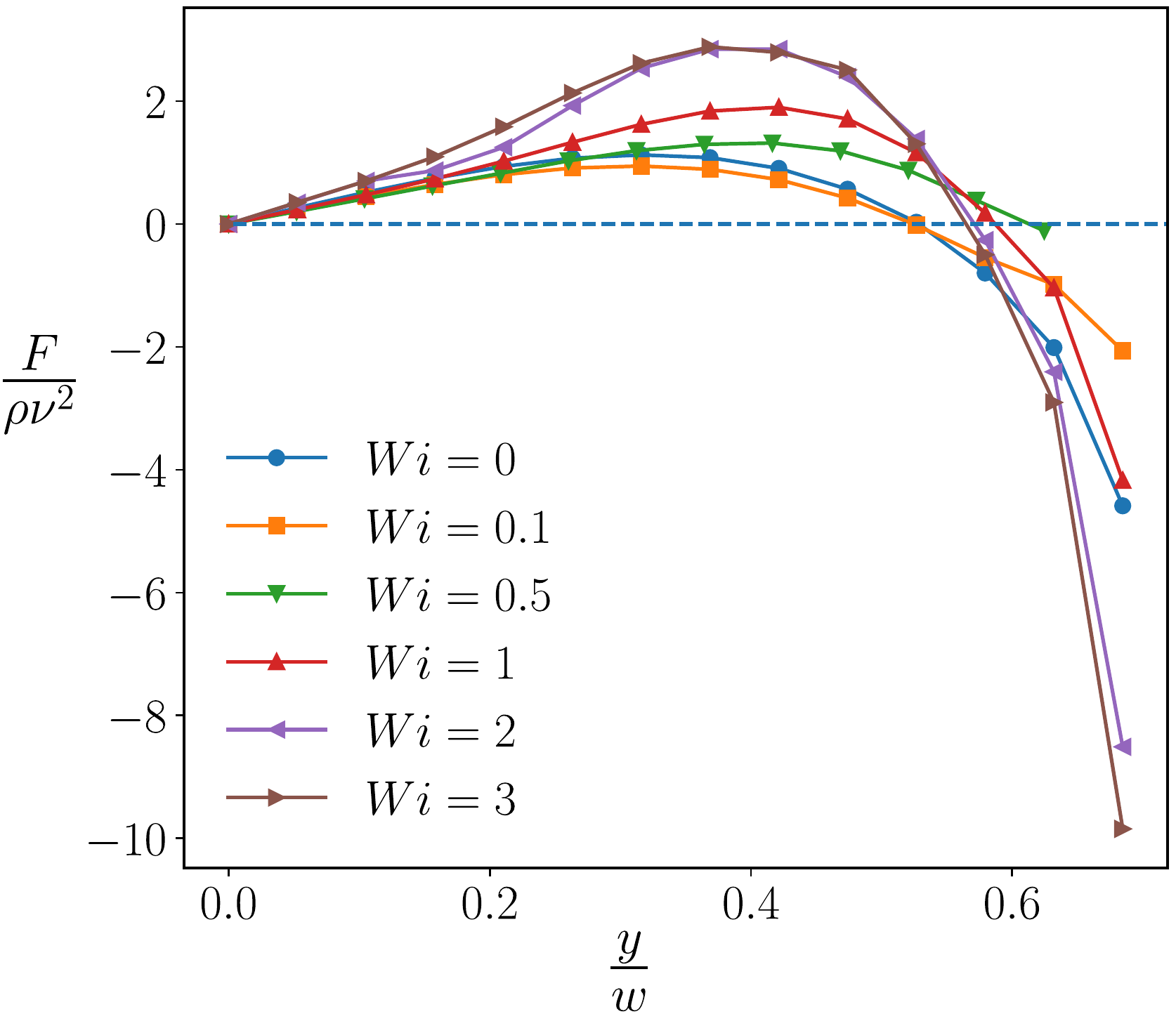}
        \caption{}
    \end{subfigure}%
    ~
    \begin{subfigure}[t]{.5\linewidth}
        \centering
        \includegraphics[height=3.1in]{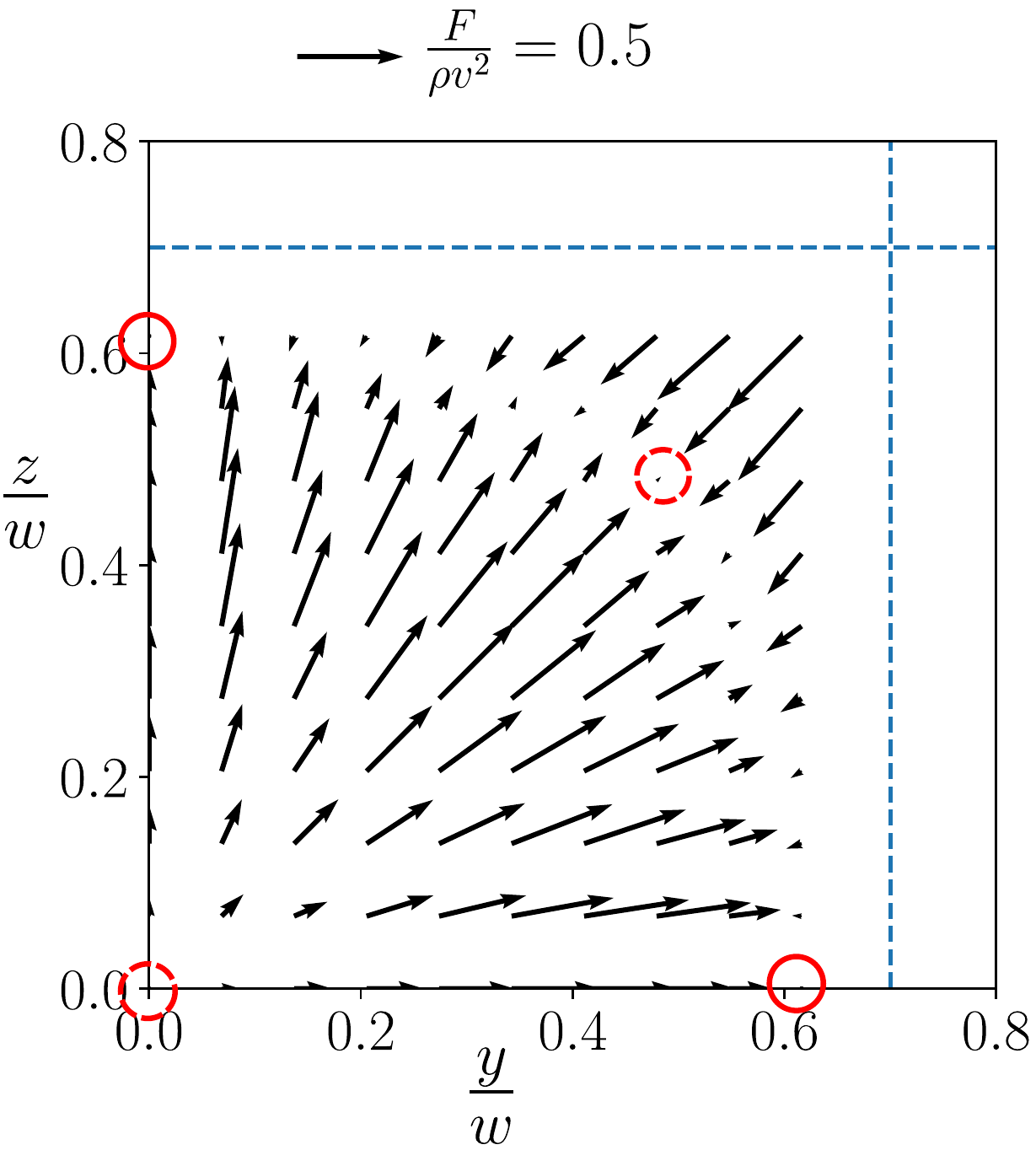}
        \caption{}
    \end{subfigure}
\caption{Force profile for (a) $Re=30$ and (b) the force-map at $Re=30$ and $Wi=0.5$.}
    \label{H_1}
\end{figure*}

With increasing the Reynolds number, the particle dynamics changes significantly from those observed in previous sections. Figure. \ref{H_1}(a) shows the force profile for $Re=30$. The force has a similar trend along the channel main axis for the entire range of $Wi$ number. The force has a positive value near the center and it becomes negative in the near wall region, i.e., the particle is pushed away from the center and the wall, which leads to the existence of an off-center equilibrium point along the main axis. As opposed to previous sections, where the elasticity changes the force profile significantly, the force distribution across the microchannel remains unchanged for the entire range of $Wi$ number studied in this work.
The slope of the force profile at the center and the off-center equilibrium points indicates that the particle is unstable and stable at these points, respectively. According to the force-map shown in Fig. \ref{H_1}(b), there is an unstable equilibrium point along the diagonal of the channel at $Re=30$ and $Wi=0.5$ that is similar to that of a Newtonian fluid ($Wi=0$). 
Hence, the behavior discovered in the numerical and experimental studies by \cite{seo2014lateral} and \cite{li2015dynamics} can be rationalized using our computational results. As shown in Fig. \ref{H_1}(a), the force gets stronger along most of the main axis as the $Wi$ number increases (for $Wi>0.1$). Hence, the particles is pushed away from the center and the wall faster, increasing the transverse migration  toward the annulus ring and leading to a smaller critical length of microchannel required for particle focusing. However, this trend is not observed for $Wi=0.1$ as the force magnitude is smaller than that of a Newtonian fluid.   

\begin{figure*}[h!]
    \centering
\begin{subfigure}[t]{.5\linewidth}
        \centering
        \includegraphics[height=2.6in]{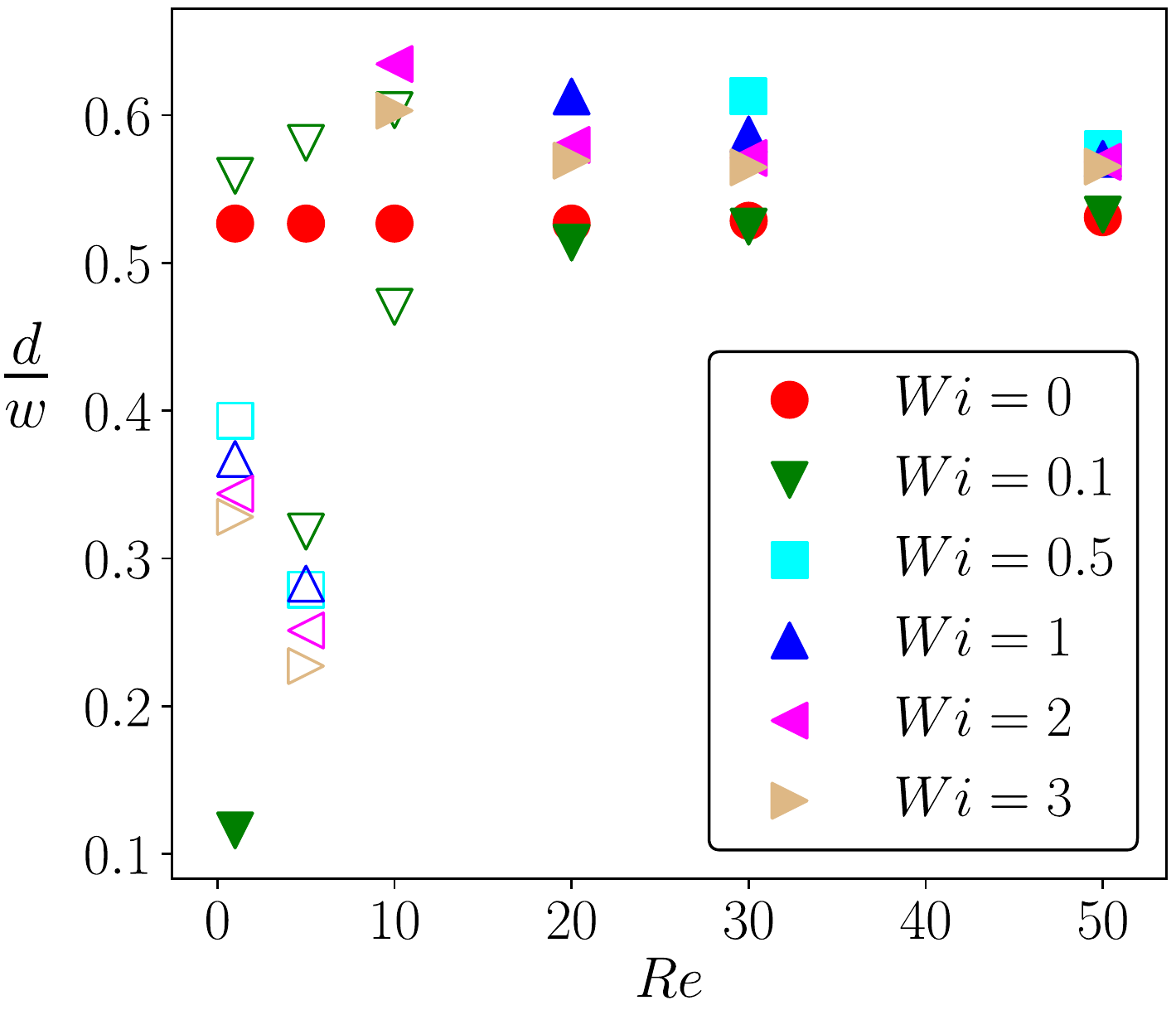}
        \caption{}
    \end{subfigure}%
    ~
    \begin{subfigure}[t]{.5\linewidth}
        \centering
        \includegraphics[height=2.6in]{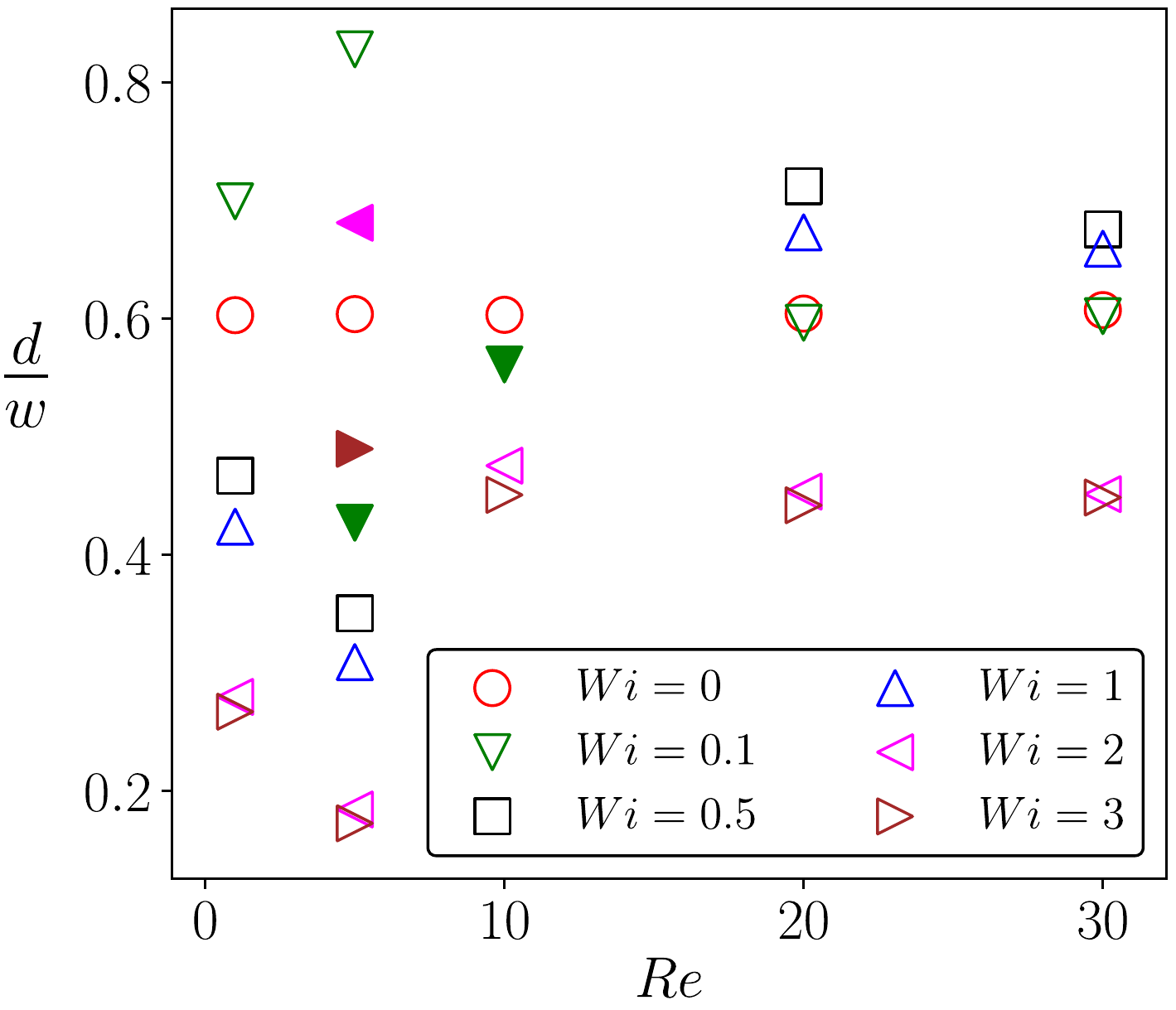}
        \caption{}
    \end{subfigure}
\caption{Distance of the off-center equilibrium points from the channel center along (a) the main axis and (b) the diagonal of the channel. Filled symbols indicate stable equilibrium points and open symbols represent  unstable equilibrium points.  }
    \label{H_2}
\end{figure*}

Figure \ref{H_2} illustrates the distance of the off-center equilibrium points from the channel center and their stability for the entire range of $Re$ and $Wi$ numbers studied in this work along the main axis and the diagonal of the channel.
As shown in Fig. \ref{H_2}(a), the equilibrium points for $Wi=0$ are all stable on the main axis and their location does not change with $Re$, which is in agreement with previous results found in Li et al. \cite{li2015dynamics}. However, the results show that the equilibrium points for $Wi\neq0$ are mostly unstable in the  low inertial regime ($Re=1,5$). The distance of these points are smaller than that of a Newtonian fluid and they approach the channel center with increasing $Wi$ number (for $Wi>0.5$), indicating reduction of trapping area with increasing the elastic effects. Oppositely, the equilibrium points for $Wi\neq0$ are all stable on the main axis in the  high inertial regime ($Re=20,30,50$). In this range of $Re$ number, the equilibrium points at $Wi=0.1$ have relatively the same distance from the channel center as that of a Newtonian fluid which is due to the small effect of elasticity compared to inertial effect, while for a higher $Wi$ number this distance is larger than that in a Newtonian fluid. The results also show that the difference between the location of equilibrium points becomes smaller as the $Re$ number increases. This can be attributed to the dominant effect of the flow inertia compared to the elastic effects. In contrary to the stability of the equilibrium points on the main axis, Fig. \ref{H_2}(b) indicates that most of equilibrium points on the diagonal of the channel are unstable. The results show that the equilibrium points approach the channel center with increasing $Wi$ number for low inertial effects, while this behavior changes for larger inertia such that the distance between equilibrium points and channel center initially increases and subsequently the equilibrium point approaches the center with increasing the $Wi$ number.
\begin{figure*}[h!]
    \centering
\begin{subfigure}[t]{.5\linewidth}
        \centering
        \includegraphics[height=2.6in,keepaspectratio]{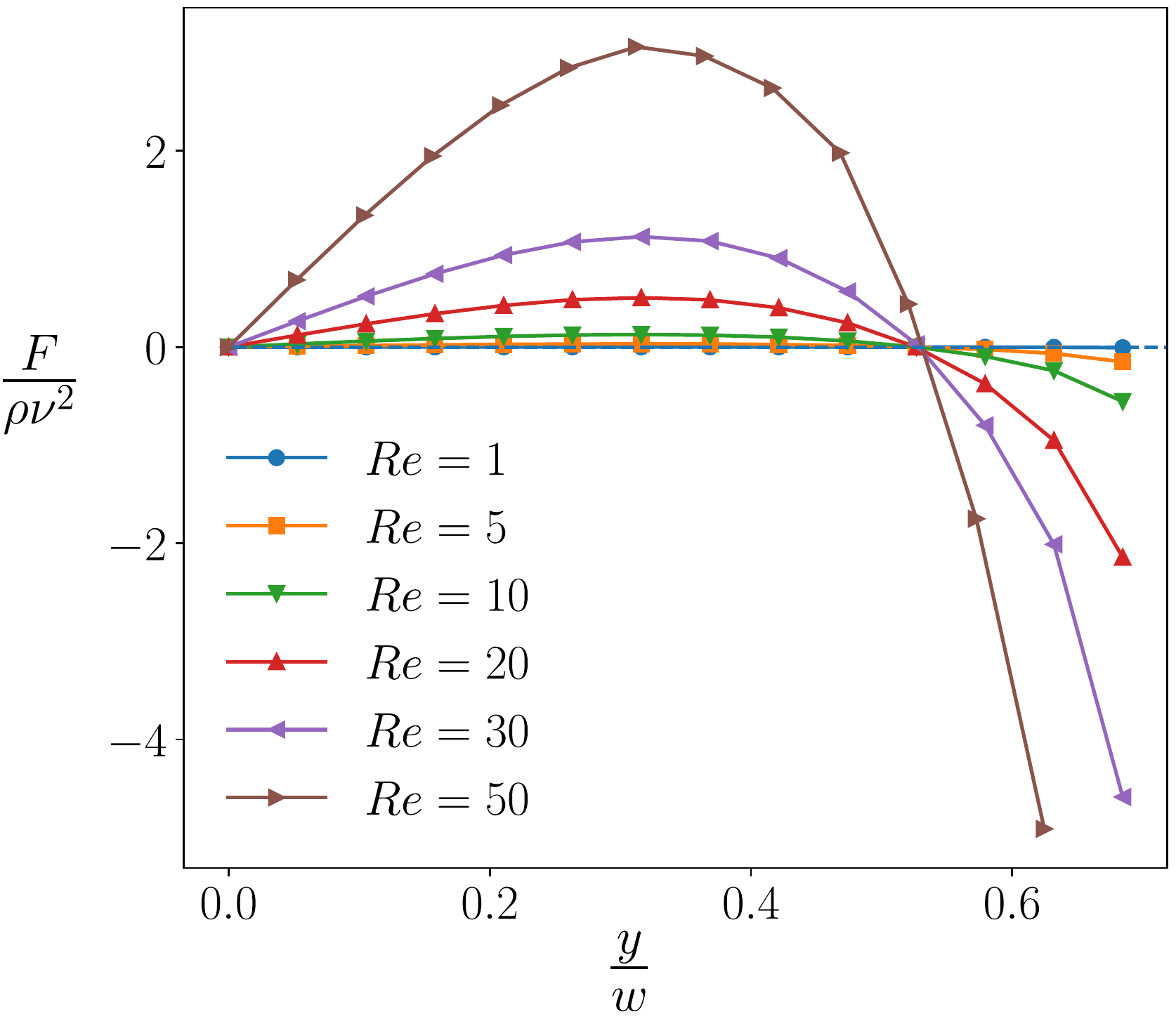}
        \caption{}
    \end{subfigure}%
    ~
    \begin{subfigure}[t]{.5\linewidth}
        \centering
        \includegraphics[height=2.6in,keepaspectratio]{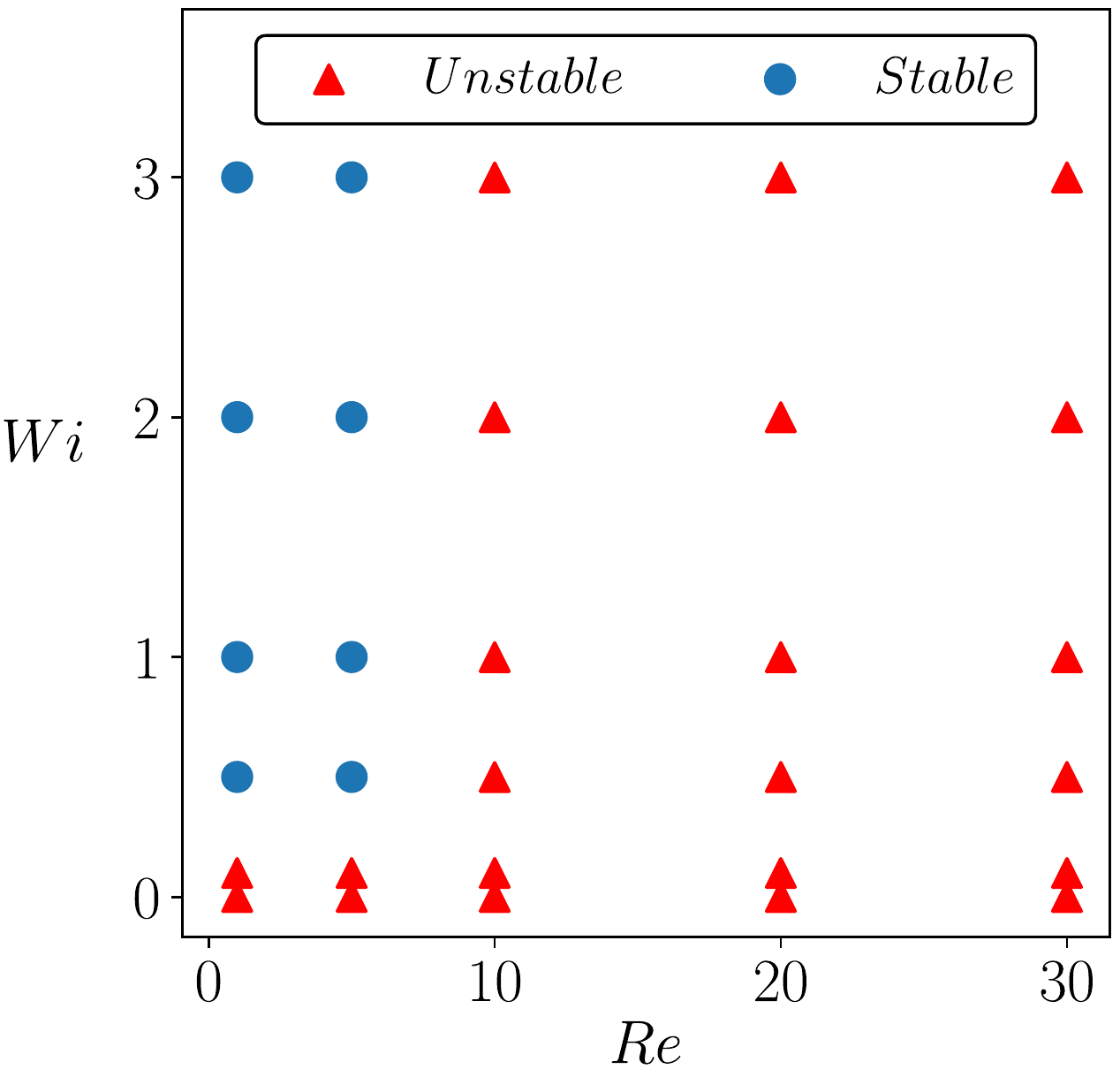}
        \caption{}
    \end{subfigure}
\caption{(a) The force profile along the main channel for $Wi=0$ and (b) the stability of channel centerline equilibrium point for a range of $Re$ and $Wi$ numbers }
    \label{H_3}
\end{figure*}
The location of equilibrium points along the main axis and the diagonal of the channel does not change with the $Re$ number in a Newtonian fluid. In order to explain this phenomenon, the force profile along the channel main axis is plotted for $Wi=0$ in Fig. \ref{H_3}(a). According to this figure, the change in the $Re$ number changes  the magnitude of the lateral force acting on the particle. As the inertial effect becomes stronger the lateral force magnitude increases significantly, however, the location at which the lateral force becomes zero is the same for the entire range of $Re$ number studied in this work. This explains the results reported by \cite{li2015dynamics} In order to investigate the combined effects of inertia and elasticity on the stability of the channel centerline equilibrium point, the stability phase diagram is illustrated in Fig. \ref{H_3}(b). As shown in this figure, the channel center is a stable equilibrium point when the elastic effect dominates the inertial effect, while with increasing the $Re$ number the focal pattern looks similar to that of a Newtonian fluid, where the particle is not expected to travel toward the channel center.

\section{Conclusions}
In this work, we conduct 3D numerical simulations to find the distribution of lift force acting on the particle in viscoelastic fluids. As a result, we predict the location of equilibrium points and their corresponding stability for a wide range of parameter space which is important  for designing  microfluidic devices relying on viscoelastic effects.
The results for low inertial regime show that  the force field acting on the particle changes significantly with increasing the elasticity of the fluid. In the case of a Newtonian fluid the particle is pushed away from the wall and the channel center to reach an annulus ring across the channel. Subsequently, the particle moves toward the equilibrium points on the main axis. Our results show that only off-center equilibrium points along the main axes are stable and other equilibrium points are unstable in a Newtonian fluid. Increasing the elasticity changes  the induced force field such that the main axis has no stable equilibrium point and the particle focuses on the corner and off-center points along the diagonal of the channel. Further increase in the Weissenberg number leads to  dominance of the elastic force over the inertial lift force, shifting the off-center equilibrium points on the diagonal of the channel toward the channel center. 

In the intermediate inertial regime, we also observe various focal patterns  by changing the fluid elasticity.  In the low $Wi$ number range, the particle focuses at the wall face center and off-center points along the diagonal of the channel. This configuration changes for larger $Wi$ numbers, where the particle is driven radially toward the corner in the entire channel cross-section, leading the corners to be the only basin of attraction in the microchannel. Further increase in the elasticity results in a configuration similar to that of a Newtonian fluid, in which the particle aggregates only at an off-center point on the main axis and equilibrium points on the center and diagonal of the channel are unstable. By splitting the total lift force experienced by the particle into inertial and elastic lift forces we conclude that  the elastic effect modifies the velocity field in the microchannel. As a result, the inertial lift force changes accordingly, leading to various particle configurations. However, the direction and the magnitude of the elastic force remain relatively unchanged in the intermediate inertial regime implying that the  fluid elasticity affects the particle dynamics indirectly by changing the velocity field. Our results for the high inertial flow indicate that the force profiles are similar for the entire range of $Wi$ number studied in this work. Accordingly the particle configuration is also similar to that of a Newtonian fluid, in which the particles aggregate only at the off-center point along the main axis.

\section{Acknowledgement}
This research was partially supported by a Grant from National Science Foundation [CBET-1705371]. This work used the Extreme Science and Engineering Discovery Environment (XSEDE) \cite{XSEDE}, which is supported by the National Science Foundation grant number ACI-1548562 through allocation TG-CTS180066.




\bibliography{achemso-demo}

\end{document}